\documentclass[iop,apj]{emulateapj}

\usepackage{acronym}
\usepackage{booktabs}
\usepackage{dcolumn}
\usepackage{graphicx}
\usepackage{amsmath,amsfonts,amssymb}
\usepackage{mathrsfs}
\usepackage[utf8]{inputenc}
\usepackage[normalem]{ulem}

\usepackage[breaklinks,colorlinks,citecolor=blue]{hyperref}

\newcolumntype{d}[1]{D{.}{.}{#1}}

\usepackage{natbib}
\usepackage{color}

\usepackage{graphicx,psfrag}
\usepackage{mathrsfs}
\usepackage{amsmath,amsfonts,amssymb}
\usepackage{multirow}
\usepackage{comment}
\usepackage{ulem}
\usepackage{hyperref}

\newcommand{\be}{\begin{equation}}
\newcommand{\ee}{\end{equation}}
\newcommand{\bea}{\begin{eqnarray}}
\newcommand{\eea}{\end{eqnarray}}
\newcommand{\bel}{\begin{align}}
\newcommand{\eel}{\end{align}}

\newcommand{\wind}{spiral-wave wind}

\def\p{\partial}

\def\i{{\rm i}}

\def\Msun{{\rm M_{\odot}}}
\def\GMc2{{\rm G M_{\odot} c^{-2}}}

\usepackage{pifont} 

\usepackage{color}
\definecolor{cyan}{rgb}{0,0.9,0.9}
\definecolor{orange}{rgb}{0.9,0.5,0}
\definecolor{magenta}{rgb}{1,0,1}
\definecolor{purple}{rgb}{0.8,0.4,0.8}
\definecolor{gray}{rgb}{0.8242,0.8242,0.8242}

\definecolor{cadmiumgreen}{rgb}{0.0, 0.42, 0.24}

\newacro{ADM}{Arnowitt-Deser-Misner}
\newacro{AMR}{adaptive mesh-refinement}
\newacro{BH}{black hole}
\newacro{BBH}{binary black-hole}
\newacro{BHNS}{black-hole neutron-star}
\newacro{BNS}{binary neutron star}
\newacro{CCSN}{core-collapse supernova}
\newacroplural{CCSN}[CCSNe]{core-collapse supernovae}
\newacro{CMA}{consistent multi-fluid advection}
\newacro{DG}{discontinuous Galerkin}
\newacro{HMNS}{hypermassive neutron star}
\newacro{EM}{electromagnetic}
\newacro{ET}{Einstein Telescope}
\newacro{EOB}{effective-one-body}
\newacro{EOS}{equation of state}
\newacroplural{EOS}[EOSs]{equations of state}
\newacro{FF}{fitting factor}
\newacro{GR}{general relativity}
\newacro{GRLES}{general-relativistic large-eddy simulation}
\newacro{GRHD}{general-relativistic hydrodynamics}
\newacro{GRMHD}{general-relativistic magnetohydrodynamics}
\newacro{GW}{gravitational wave}
\newacro{ILES}{implicit large-eddy simulations}
\newacro{LIA}{linear interaction analysis}
\newacro{LES}{large-eddy simulation}
\newacroplural{LES}[LES]{large-eddy simulations}
\newacro{MRI}{magnetorotational instability}
\newacro{NR}{numerical relativity}
\newacro{NS}{neutron star}
\newacro{PNS}{protoneutron star}
\newacro{SASI}{standing accretion shock instability}
\newacro{SGRB}{short $\gamma$-ray burst}
\newacro{SN}{supernova}
\newacroplural{SN}[SNe]{supernovae}
\newacro{SNR}{signal-to-noise ratio}

\begin{document}

\title{Spiral-wave wind for the blue kilonova}

\author{Vsevolod Nedora\altaffilmark{1},
  Sebastiano Bernuzzi\altaffilmark{1},
  David Radice\altaffilmark{2,3,4},
  Albino Perego\altaffilmark{5,6},
  Andrea Endrizzi\altaffilmark{1},
  N\'estor Ortiz\altaffilmark{1,7}}

\altaffiltext{1}{Theoretisch-Physikalisches Institut, Friedrich-Schiller-Universit{\"a}t Jena, 07743, Jena, Germany}
\altaffiltext{2}{Institute for Advanced Study, 1 Einstein Drive,
Princeton, NJ 08540, USA}
\altaffiltext{3}{Department of Astrophysical Sciences, Princeton University,
4 Ivy Lane, Princeton, NJ 08544, USA}
\altaffiltext{4}{Department of Physics, The Pennsylvania State University, University Park, PA 16802, USA}
\altaffiltext{5}{Dipartimento di Fisica, Universit\'a di Trento, Via Sommarive 14, 38123 Trento, Italy}
\altaffiltext{6}{Istituto Nazionale di Fisica Nucleare, Sezione di Milano-Bicocca, Piazza della Scienza 20100, Milano, Italy}
\altaffiltext{7}{Instituto de Ciencias Nucleares, Universidad Nacional Aut\'onoma de M\'exico,
Circuito Exterior C.U., A.P. 70-543, M\'exico D.F. 04510, M\'exico}

\begin{abstract}
  The AT2017gfo kilonova counterpart of the binary neutron
  star merger event GW170817 was characterized by an
  early-time bright peak in optical and UV bands. Such
  blue kilonova is commonly interpreted as a signature of  
  weak $r$-process nucleosynthesis
  in a fast expanding wind whose origin is currently debated.
  Numerical-relativity simulations with 
  microphysical equations of state, approximate neutrino transport, and turbulent
  viscosity reveal a new hydrodynamics-driven 
  mechanism 
  that can power the blue kilonova.
  Spiral density waves in the remnant
  generate a characteristic wind 
  of mass ${\sim}10^{-2}~\Msun$ and velocity ${\sim}0.2$c.
  The ejected material has electron fraction mostly distributed above $0.25$ being partially
  reprocessed by hydrodynamic shocks in the expanding arms. The
  combination of dynamical
  ejecta and \wind{} can account for solar system abundances of $r$-process
  elements and early-time observed light curves.
\end{abstract}


\section{Introduction}
\label{sec:intro}

The observation of the kilonova (kN) AT2017gfo \citep{Coulter:2017wya,Chornock:2017sdf,Nicholl:2017ahq,Cowperthwaite:2017dyu,Tanvir:2017pws,Tanaka:2017qxj}
associated to the binary neutron star (BNS) merger GW170817 \citep{TheLIGOScientific:2017qsa} provided
evidence 
that the ejection of neutron-rich matter from compact binary mergers is a primary site for
$r$-process
nucleosynthesis~\citep{Lattimer:1974a,Li:1998bw,Kulkarni:2005jw,Rosswog:2005su,Metzger:2010sy,Roberts:2011xz,Kasen:2013xka}.
In this scenario, the electromagnetic UV/optical/NIR transient is
powered by the radioactive decay of the freshly synthesized elements.
The NIR luminosity of AT2017gfo peaked at several days after the
merger \citep{Chornock:2017sdf}, and it is consistent with expectations
that the opacities of expanding $r$-process material are dominated by the opacities of lanthanides and possibly actinides \citep{Kasen:2013xka}. 
The UV/optical luminosity peaked instead in less than one day after the
merger \citep{Nicholl:2017ahq}, and it
originates from ejected material that experienced only a partial
$r$-process nucleosynthesis \citep{Martin:2015hxa}.
A fit of AT2017gfo light curves to a semianalytical two-components 
spherical model indicates a lanthanide poor (rich) blue (red) component of mass
$2.5\times10^{-2}M_{\odot}$ ($5.0\times10^{-2}M_{\odot}$) and velocity
$0.27$c ($0.15$c)
\citep{Cowperthwaite:2017dyu,Villar:2017wcc}
(See however \citep{Waxman:2017sqv} for an alternative interpretation.)
Similar results are obtained using more sophisticated 1D simulations of
radiation transport along spherical shells of mass ejecta
\citep{Tanvir:2017pws,Tanaka:2017qxj}.

Numerical relativity (NR) simulations produce dynamical ejecta of a few
times $10^{-3}\Msun$  
with velocities distributed around ${\sim} 0.1{-}0.3$c
~\citep{Hotokezaka:2012ze,Bauswein:2013yna,Radice:2018pdn}.
Dynamical ejecta are characterized by a range of electron fractions 
$0.05\lesssim Y_e \lesssim 0.4$;
with larger values distributed towards polar regions above the
remnant (as part of the shocked component) and lower values across the equatorial plane.
These properties are largely independent of the NS equation of state (EOS)~\citep{Sekiguchi:2015dma,Radice:2018pdn}.
Additional ejecta from the disk are expected on longer
timescales~\citep{Perego:2014fma,Just:2014fka,Kasen:2014toa,Metzger:2014ila,Wu:2016pnw,Siegel:2017nub,Fujibayashi:2017puw,Miller:2019dpt}; 
disk mass and composition depend on the
binary mass and EOS \citep{Radice:2017lry,Perego:2019adq}.
Neutrino irradiation can unbind ${\sim}5$\% of the disk mass with
$Y_e>0.25$ and velocities ${\lesssim}0.08$c from the polar
region~\citep{Perego:2014fma,Martin:2015hxa}.
A significant fraction of the disk mass, up to 40\%, can be
ejected on time scales ${\gtrsim}100$~ms due to magnetic-field induced viscosity and/or nuclear recombination,~\citep{Dessart:2008zd,Fernandez:2014bra,Wu:2016pnw,Lippuner:2017bfm,Siegel:2017nub,Fujibayashi:2017puw,Radice:2018xqa,Fernandez:2018kax,Miller:2019dpt}.
These secular ejecta are expected to have velocities ${\lesssim}0.05{-}0.1$c
and electron fraction in the broad range 
$0.1\lesssim Y_e \lesssim 0.5$, where
lower (higher) values are found for black-hole (long-lived NS) remnant.
If present, the secular ejecta might give the dominant contribution to the
kN on timescales of days to months \citep{Fahlman:2018llv}.

KN light curve models need
to account for multiple ejecta (dynamical, wind,
viscous, etc.), for the anisotropy of the ejecta composition, and
for the irradiation among the ejecta components to fully explain AT2017gfo.
Indeed, outflow properties inferred for AT2017gfo using multi-components and 2D kN models including these effects are broadly compatible with the results from simulations, e.g. \citep{Perego:2017wtu,Kawaguchi:2018ptg}.
The early blue kN however, remains a challenging
aspect to model. Both semi-analytical and radiation transport
models require ejecta properties different from those found in
simulations. In particular, simulations cannot produce ejecta with the
large velocities and electron fraction inferred from the
electromagnetic data~\citep{Fahlman:2018llv}. 

There exist alternative explanations of the blue kN based
on the interaction between a relativistic jet and the ejecta
\citep{Lazzati:2016yxl,Bromberg:2017crh,Piro:2017ayh} but simulations show that successful jets do not deposit a sufficient amount of thermal energy in the ejecta for this mechanism to work \citep{Duffell:2018iig}. 
Other possibilities include the presence of highly magnetized winds \citep{Metzger:2018uni,Fernandez:2018kax},
or the presence of the so-called viscous-dynamical ejecta \citep{Radice:2018ghv}.
However, both models rely on the  development of large-scale strong magnetic fields.
Here, we identify a new generic hydrodynamics-driven mechanism that works in self-consistent
ab-initio simulations and does not require the presence of a strong ordered magnetic field. 


\section{Method}
\label{sec:}

\begin{figure}[t]
	\centering
	\includegraphics[width=0.49\textwidth]{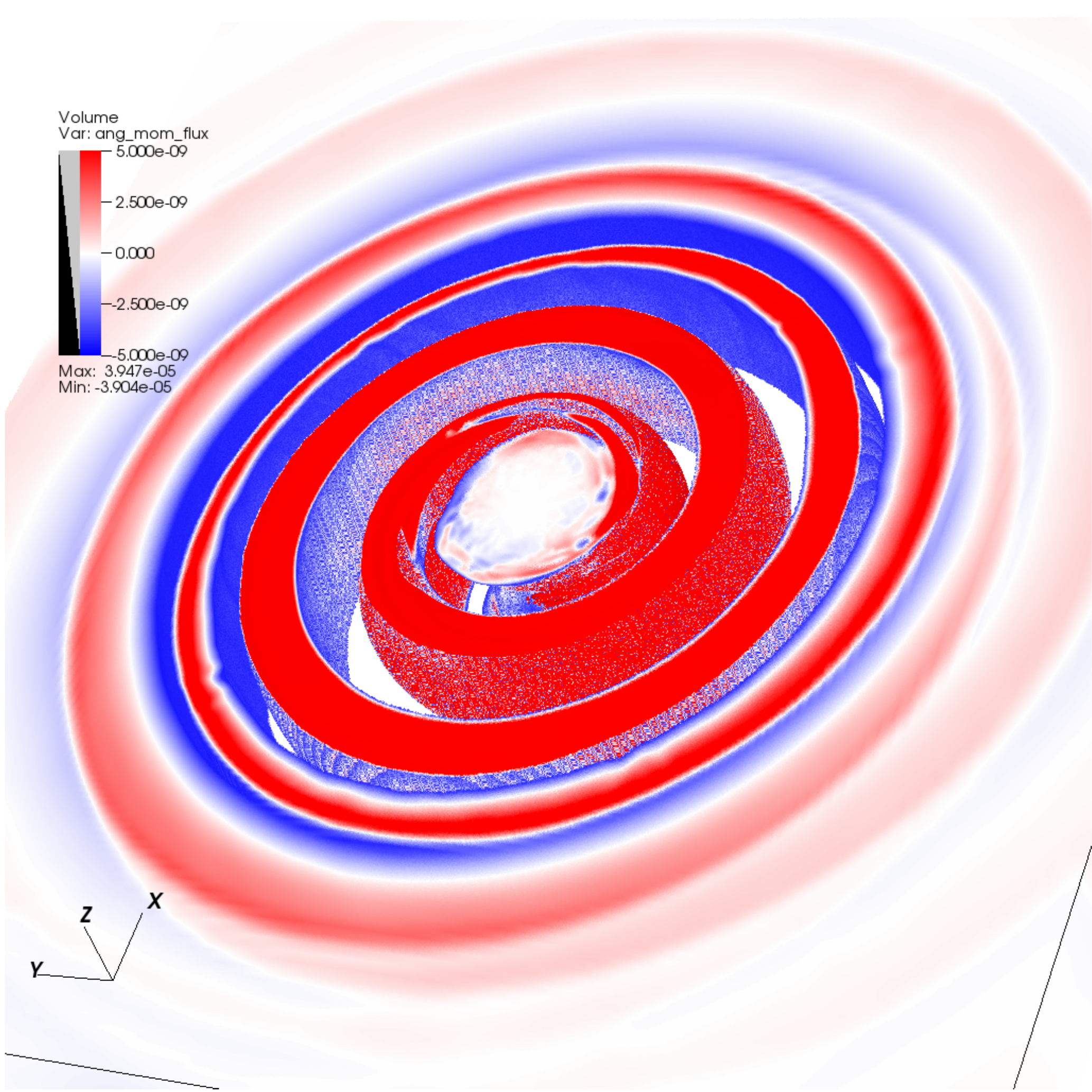}
	\caption{3D distribution of angular momentum density flux $J_r$
          from the DD2 simulation with turbulent viscosity at ${\sim}43.5$~ms after
          merger. $J_r$ is shown on a central region of
          $(89\times89\times60)$~km${}^3$ covering the remnant NS
          and disk, and it is given in units where $c=G=\Msun=1$.}
	\label{fig:ang_mom_flux}
\end{figure}

We perform 3+1 NR simulations of two binaries with mass
$M=(1.364+1.364)\Msun$ and NS described by the microphysical EOS
HS(DD2) \citep{Typel:2009sy,Hempel:2009mc} and LS220 \citep{Lattimer:1991nc}.
The simulations include the merger and the remnant evolution for a
timescale of at least 30~ms and up to 100~ms depending on the binary.
The results presented here are 
representative cases producing a long-lived NS 
remnant (DD2) and a short-lived NS (LS220) 
from a larger set of simulations that will be presented elsewhere.

We use the \texttt{WhiskyTHC} code
\citep{Radice:2012cu,Radice:2013hxh,Radice:2013xpa,Radice:2018xqa}
with the approximate neutrino
transport scheme developed in \citep{Radice:2016dwd,Radice:2018pdn}.
The simulations treat turbulent viscosity using
the general-relativistic large eddy 
simulations method (GRLES) \citep{Radice:2017zta}.
The interactions between the fluid and neutrinos are treated with a 
leakage scheme in the optically thick regions
\citep{Ruffert:1995fs,Neilsen:2014hha} while free-streaming neutrinos
are evolved according to the M0 scheme discussed in Ref. \citep{Radice:2018pdn}. 
The turbulent viscosity in the GRLES is parametrized as
$\sigma_T = \ell_{\rm mix} c_s$, where $c_s$
is the sound speed and $\ell_{\rm mix}$  is a free parameter 
that depends on the intensity of the turbulence. We perform two groups of simulations in this work with 
$\sigma_T$ either set to zero, or prescribed as a function of the 
rest-mass density as in \citep{Perego:2019adq} using the
results of \citep{Kiuchi:2017zzg}. 
We perform simulations with the same grid setup as in Ref. \citep{Radice:2018pdn}.
In particular, the adaptive mesh refinement grids have seven
2:1 refinement levels with finest linear resolutions of $h=246, 185,
123$m, which are labelled LR, SR and HR.
  Each model was evolved at least at two different resolutions (LR and SR).

The ejecta are calculated on coordinate spheres at $r=294$~km
employing the geodesic criterion for the dynamical
ejecta~\citep{Radice:2018pdn}. For the wind we use the Bernoulli
criterion, which is appropriate for steady-state flow, assuming
$(\p_t)^a$ is an approximate Killing vector (see
e.g.~\citep{Kastaun:2014fna}). The Bernoulli calculation is started
after the ejecta mass computed with the geodesic criterion has
saturated to its final value. 
From the fluid's stress energy tensor,
we compute the angular momentum density flux $J_r = T_{ra}(\p_\phi)^a$,
where $\phi$ is the cylindrical angular coordinate;
angular momentum is conserved if $(\p_\phi)^a$ is a Killing vector.
$r$-process nucleosynthesis yields are computed using the method
detailed in \citep{Radice:2018pdn}.


\section{Results}
\label{sec:}

\begin{figure}[t]
	\centering
	\includegraphics[width=0.49\textwidth]{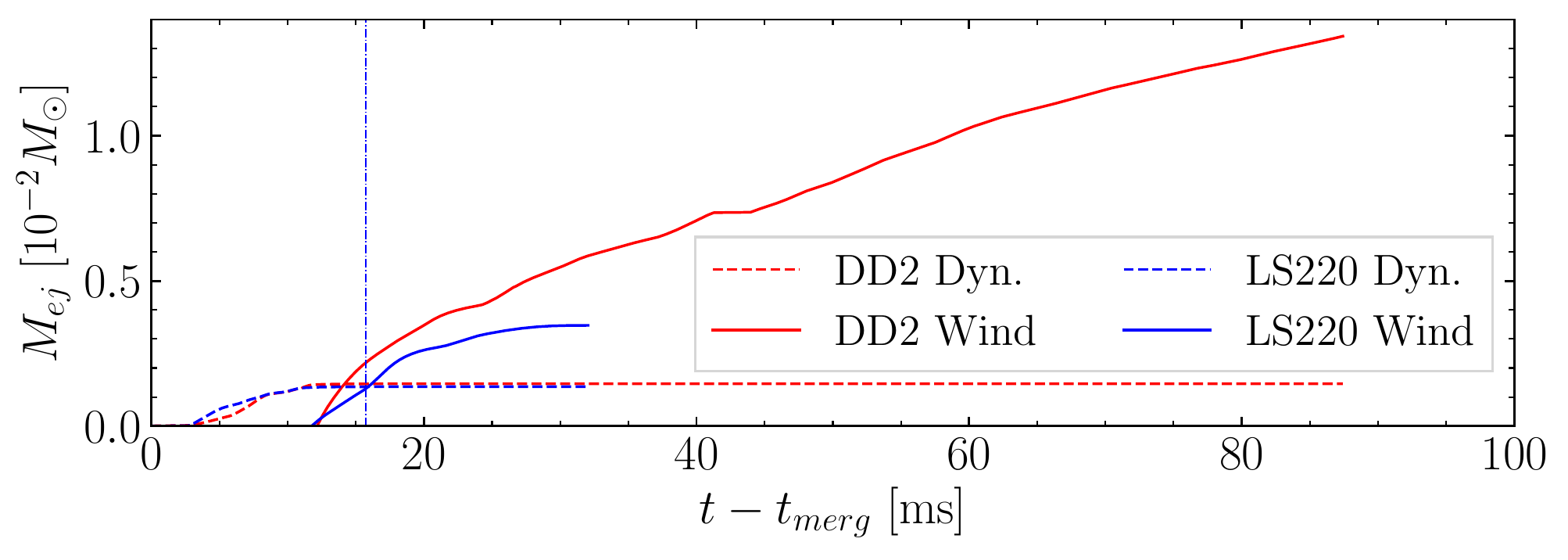}
	\includegraphics[width=0.49\textwidth]{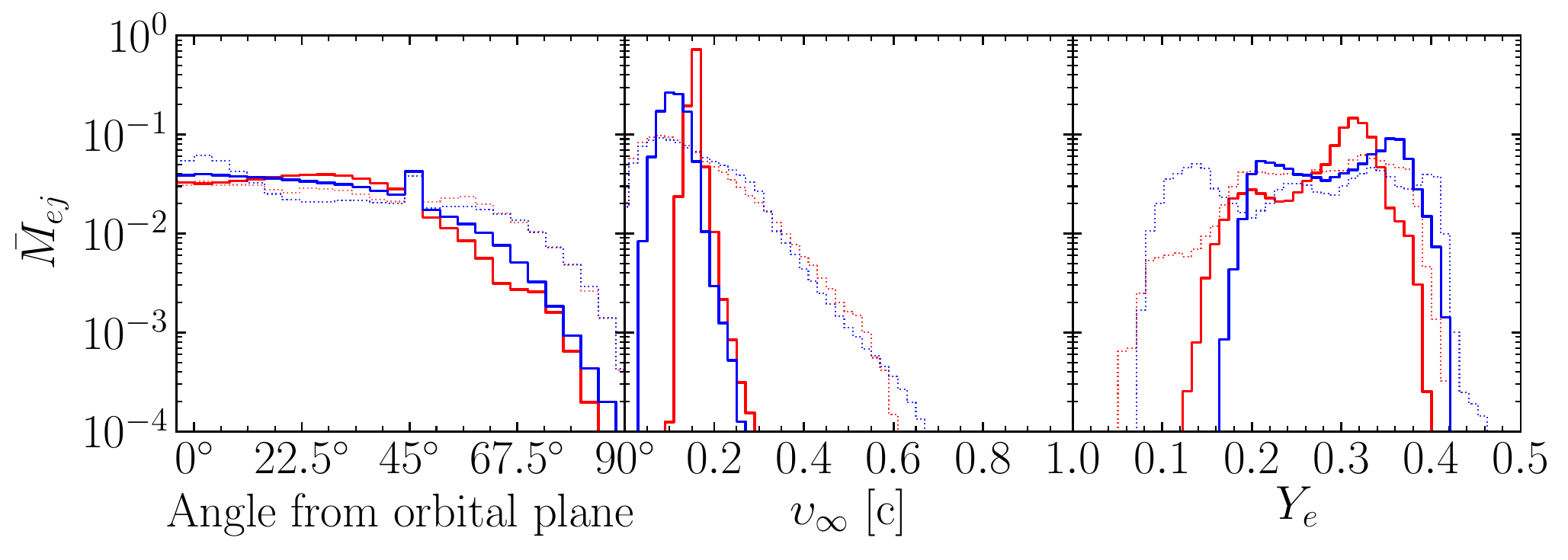}
	\includegraphics[width=0.49\textwidth]{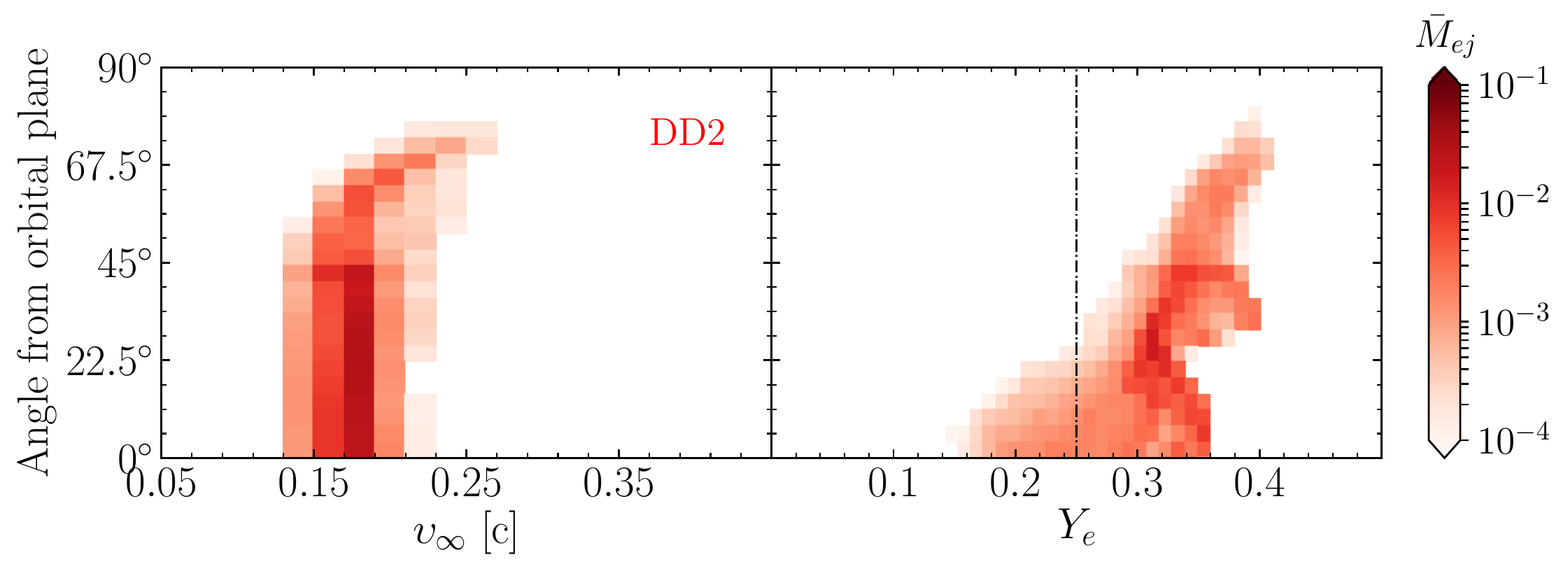}
	\caption{Properties of the \wind{} and dynamical ejecta
          computed form the simulations with turbulent viscosity.
          Top: evolution of unbound mass for dynamical ejecta
          (dashed lines)
          and \wind{}
          (solid lines). $t=0$ marks the moment of merger, the vertical
          line marks the collapse time of the LS220 BNS.
          Middle: mass histograms for the angular (left), velocity (center) and electron
          fraction (right) distributions.
          Bottom: angular distribution and composition of the \wind{}
          for DD2.
          Note the $\bar{M}_{ej}$ in the middle and bottom panels is normalized to one.}
	\label{fig:ej_properties}

\end{figure}

The key dynamical feature of relevance here is the
development of spiral arms in the remnant
\citep{Shibata:1999wm,Shibata:2006nm,Bernuzzi:2013rza,Kastaun:2014fna,Bernuzzi:2015opx,East:2015vix,Paschalidis:2015mla,Radice:2016gym,Lehner:2016wjg}.
The hydrodynamic instability is monitored by a decomposition in Fourier modes
$e^{-\i m\phi}$ of the Eulerian rest-mass density on the equatorial plane 
[see Eq.~(1) of \citep{Radice:2016gym}] and characterized by the
development of a $m=2$ followed by a $m=1$ mode 
\citep{East:2015vix,Paschalidis:2015mla,Radice:2016gym,Lehner:2016wjg,Bernuzzi:2013rza,Kastaun:2014fna}.
In the short-lived remnant (LS220) the $m=1$ mode
is subdominant with respect to the $m=2$, and it reaches a maximum close to the collapse
\citep{Bernuzzi:2013rza}. Instead, in the long-lived remnant (DD2) the $m=1$
becomes the dominant mode at $\sim$20~ms and persists throughout the
remnant's lifetime, while the $m=2$ efficiently dissipates via
gravitational-wave emission \citep{Bernuzzi:2015opx,Radice:2016gym}.
Considering the turbulent viscosity effect, we find that
the $m=2$ mode is suppressed more rapidly in presence of viscosity than
without viscosity. By contrast, the $m=1$ modes are not
significantly affected by viscosity.
The spiral arms propagate from the remnant NS into
the disk and transport angular momentum outwards as shown in
Fig.~\ref{fig:ang_mom_flux}. Such global density waves are a generic and
efficient mechanism to redistribute energy and eventually deplete  
accretion disks \citep{Goodman:2001a,Rafikov:2016a,Arzamasskiy:2018a}. 
Crucially, we find that both the $m=1$ and $m=2$ modes generate a \wind{} from the
disk's outer layers that is distinct from the dynamical ejecta, see Fig.~\ref{fig:ej_properties}.

The long lived NS remnant (DD2) develops a \wind{} more massive than the dynamical ejecta,
as shown also in Fig.~\ref{fig:ej_properties}.
The \wind{} mass is larger the longer the remnant survives and 
the more massive the disks are. It continues as long as as the 
remnant does not collapse and the spiral modes persist.
Thus, binary mass asymetry can enhance the \wind{}
as we find in simulations discussed elsewhere [In Prep.].
The inclusion of turbulent viscosity alters all the ejecta masses with an
additional component and, for the viscosity parametrization we
have considered, it enhances the DD2 \wind{} mass by ${\sim}25$\%.
The viscosity effect is larger than resolution effects.
Comparing data at different grid resolutions we find that the
largest variation is in the wind mass. The relative variation
of mass from data pairs at increasing resolutions is ${\sim}+15\%$
(LR-SR) and ${\sim}+8\%$ (SR-HR). Hence, finite grid effects tend to
increase mass. A similar analysis on the average electronfraction
and velocity indicate variations below $4\%$.

The \wind{} has an angular distribution of mass similar to the
dynamical ejecta with material mostly confined to the orbital plane,
as shown by the histograms in Fig.~\ref{fig:ej_properties}.
On the contrary, the velocity profiles show a drastic difference
between the two ejecta components. While the dynamical ejecta has a
broad velocity distribution \citep{Hotokezaka:2012ze,Bauswein:2013yna,Radice:2018pdn}, the 
\wind{} velocity is narrowly distributed around  
$0.2$c in the case of a long-lived remnant (DD2).
The \wind{} from the short-lived
remnant (LS220) has a broader velocity distribution extending down to
$0.1$c.
This is due to the spiral-wave shutting down and the disk
  transition to a more steady accretion. As a consequence, the \wind{}
  ceases but ejecta continue as a slower disc wind driven
  by nuclear recombination solely.
The electron fraction of the \wind{} has a
  narrower distribution than the dynamical ejecta in both
  cases. But because disks around NS remnants are less compact, 
  colder, and optically thicker than those around black
  holes~\citep{Perego:2019adq}, the outer layers of the DD2 disk have a 
  lower $Y_e$ than the LS220 disk and so does the \wind{} coming from
  those layers.
While the \wind{} is generic in its hydrodynamics origin, the
  quantification of its properties relies on the accurate microphysics and
  neutrino treatment in our simulations.
  
\begin{figure}[t]
  \centering
  \includegraphics[width=0.49\textwidth]{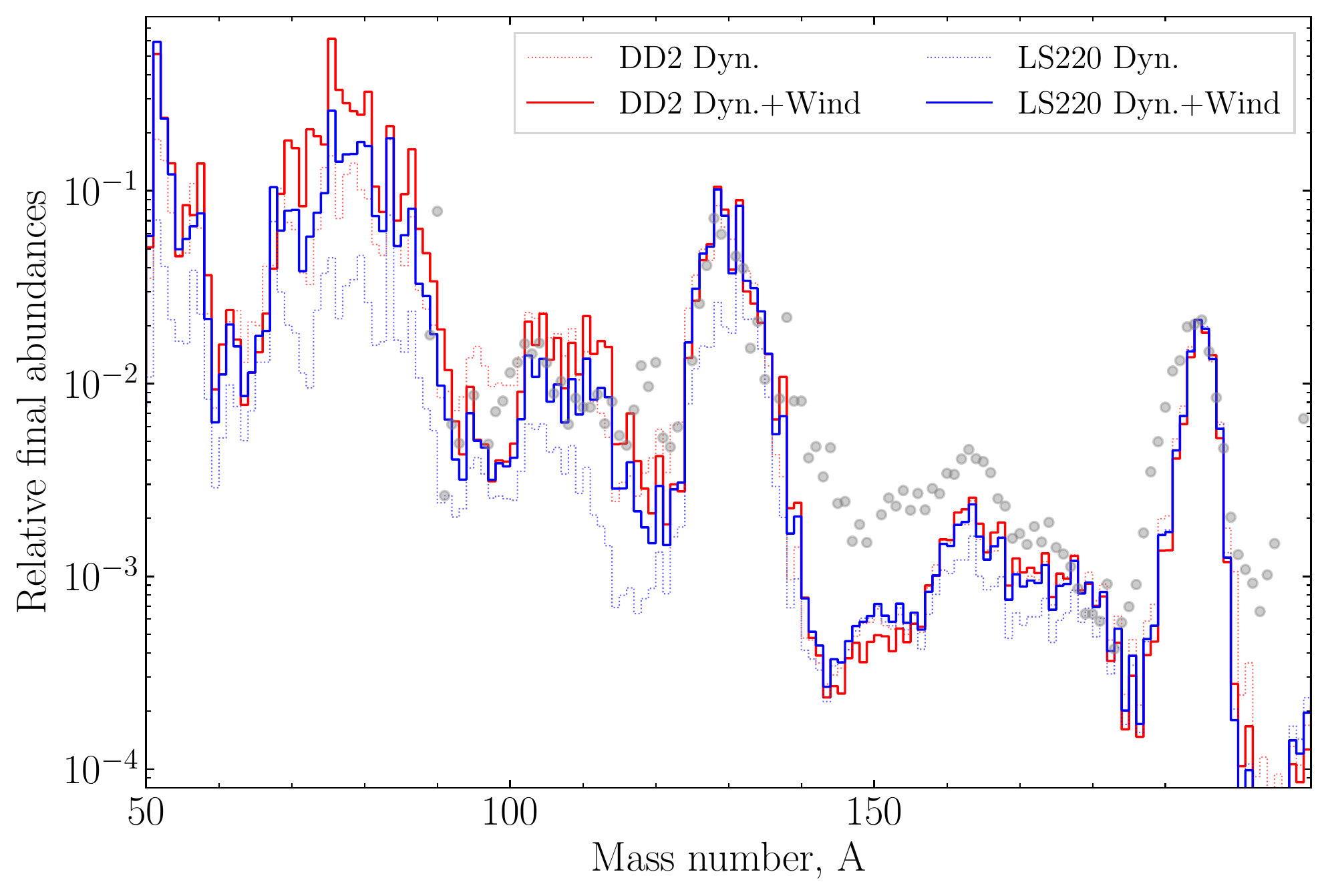}
  \caption{Nucleosynthetic yields in the ejecta. Dashed lines
    correspond to the dynamical ejecta, while solid lines are the
    summed yields including the \wind{}. Model abundances are normalized to
    $A=195$ element. Gray dots show the solar abundances from \cite{Arlandini:1999an}.}
    \label{fig:nucleo}
\end{figure}

Matter in the \wind{} undergoes $r$-process nucleosynthesis, 
and produces predominantly elements up to the second peak
(mass number $A<130$), see Fig.~\ref{fig:nucleo}. 
The combined nucleosynthesis in the dynamical ejecta and
the \wind{} reproduces the solar abundances to within the uncertainties due to nuclear physics.
The radioactive decay in the \wind{} contributes to a blue
day-long kN emission similar to the neutrino wind and viscous ejecta 
\citep{Perego:2014fma,Martin:2015hxa,Metzger:2014ila,Miller:2019dpt}.
But in comparison to the latter, the \wind{} is distributed closer 
to the equatorial plane, it is faster and more massive. 

We calculate light curves in different photometric
bands by postprocessing the simulation data with the anisotropic
multi-component model of \citep{Perego:2017wtu}.
In order to emulate the \wind{} from different BNS, 
the DD2 \wind{} data are extracted every $10$~ms until the end of the
simulation (${\sim}90$~ms) and then linearly extrapolated to $250$~ms. 
The LS220 simulation has instead a complete ejecta, since both the
dynamical and the \wind{} have terminated at the end of our simulation.
We stress that we do not include additional ejecta components to the
ones extracted from the simulations, although we expect additional
material to be unbound due to viscous processes and nuclear
recombination on even longer timescales \citep{Radice:2018xqa}.  

\begin{figure}[t]
  \centering
  \includegraphics[width=0.49\textwidth]{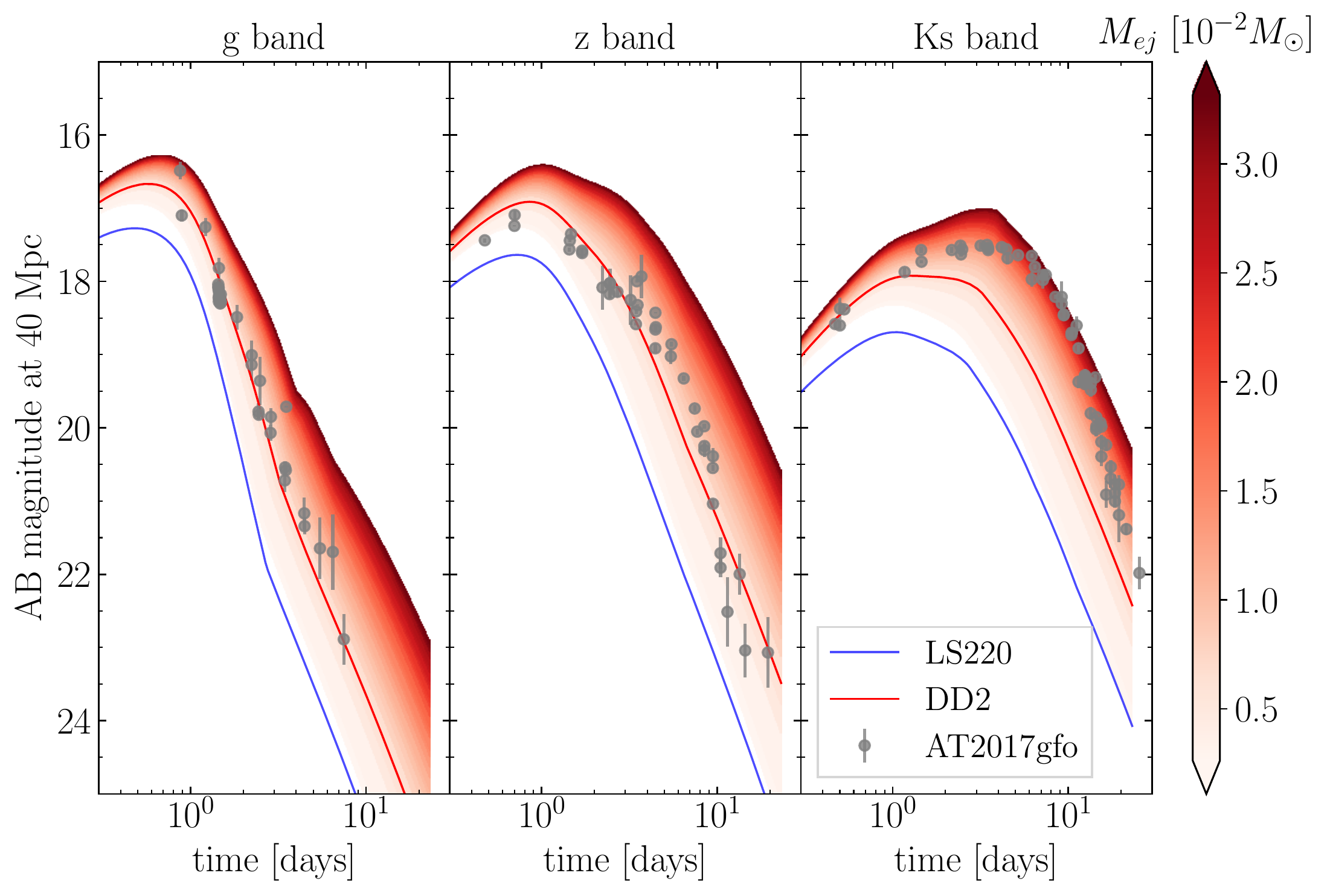}
  \caption{Bolometric kN light curves in three representative bands from blue to
    infrared for the two simulations with turbulence viscosity compared to
    AT2017gfo data from~\citep{Villar:2017wcc}.
    The color gradient is the effect related to different
    \wind{} masses, that suggests possible variations of the light
    curves for different BNS. The band is computed by extracting the
    \wind{} mass from DD2 every $10$~ms until the end of the simulation, and
    then by linearly extrapolating the data to $250$~ms.}
    \label{fig:knlc}
\end{figure}

When comparing our results to the early emission of AT2017gfo in 
Fig.~\ref{fig:knlc}, we find good agreement between the observed 
luminosities in the high frequency bands and our kN model informed by
the DD2 simulation with \wind{} masses ${\sim}0.75{-}1.25\times10^{-2}M_{\odot}$.
By contrast, the LS220 simulation does not produce enough
ejecta to explain the observations with this light-curve model.
Explaining the low frequency bands with the DD2 data 
would require a more massive \wind{} with mass ${\gtrsim}
2\times10^{-2}M_{\odot}$, implying a remnant lifetime of
${\gtrsim}200$~ms. However, a more massive \wind{} is incompatible with
  the early emission for the considered simulations.
Late-time luminosities (peaking at $t \approx 3{-}10$~days),
could be explained by a combination of \wind{} and viscous
ejecta from the disintegration of the disk. 
These results have uncertainties related to our simplified calculation of
the kilonova light curves which is expected to be less accurate at
late times when absorption features and deviations from local
thermodynamics equilibrium become more relevant,
e.g.~\citep{Smartt:2017fuw}. Indeed, time- and energy-dependent
modeling of the photon radiation transport will be needed to model
more robustly the kN emission, and quantitatively reproduce the
observed spectra
\citep{Kasen:2017sxr,Tanaka:2017qxj,Miller:2019dpt,Bulla:2019muo}. Furthermore,
all current kilonova models suffer systematic uncertainties in
nuclear (e.g. mass models, fission fragments and $\beta$-decay
rates) and atomic (e.g. detailed wavelength dependent opacities for
$r$-process element)
physics~\citep{Eichler:2014kma,Rosswog:2016dhy,Gaigalas:2019ptx}.

\section{Conclusion}
Standard kN models applied to the early AT2017gfo light curve are in
tension with ab-initio simulations conducted so far.
While alternative interpretations have been proposed, they are either
disfavored by current simulations and observations (e.g. jets) \citep{Bromberg:2017crh,Duffell:2018iig},
or require the presence of large-scale strong magnetic 
fields which might not be formed in the postmerger
\citep{Metzger:2018uni,Fernandez:2018kax,Radice:2018ghv,Ciolfi:2019fie}. 
We identified a robust dynamical mechanism for mass ejection that
explains early-time observations without requiring any fine-tuning.
The resulting nucleosynthesis is complete and produces all
$r$-process elements in proportions similar to solar system abundances.
Methodologically, our work underlines the importance of employing
NR-informed ejecta for the fitting of light-curves.
Further work in this direction should 
include better neutrino-radiation transport and magnetohydrodynamic effects
\citep{Siegel:2017nub,Fujibayashi:2017puw,Radice:2018xqa,Radice:2018pdn,Miller:2019dpt}.

\begin{acknowledgments}
SB and NO acknowledge support by the EU H2020 under ERC Starting
Grant, no.~BinGraSp-714626.  
DR acknowledges support from a Frank and Peggy Taplin Membership at the
Institute for Advanced Study and the
Max-Planck/Princeton Center (MPPC) for Plasma Physics (NSF PHY-1804048).
Computations were performed 
on the supercomputer SuperMUC at the LRZ Munich (Gauss project
pn56zo);
on the supercomputers Bridges, Comet, and Stampede
(NSF XSEDE allocation TG-PHY160025); on NSF/NCSA Blue Waters (NSF
AWD-1811236); 
on supercomputer Marconi at CINECA (ISCRA-B project number 
HP10BMHFQQ). 
\end{acknowledgments}

\bibliography{refs20191209}

\begin{thebibliography}{76}%
\makeatletter
\providecommand \@ifxundefined [1]{%
 \@ifx{#1\undefined}
}%
\providecommand \@ifnum [1]{%
 \ifnum #1\expandafter \@firstoftwo
 \else \expandafter \@secondoftwo
 \fi
}%
\providecommand \@ifx [1]{%
 \ifx #1\expandafter \@firstoftwo
 \else \expandafter \@secondoftwo
 \fi
}%
\providecommand \natexlab [1]{#1}%
\providecommand \enquote  [1]{``#1''}%
\providecommand \bibnamefont  [1]{#1}%
\providecommand \bibfnamefont [1]{#1}%
\providecommand \citenamefont [1]{#1}%
\providecommand \href@noop [0]{\@secondoftwo}%
\providecommand \href [0]{\begingroup \@sanitize@url \@href}%
\providecommand \@href[1]{\@@startlink{#1}\@@href}%
\providecommand \@@href[1]{\endgroup#1\@@endlink}%
\providecommand \@sanitize@url [0]{\catcode `\\12\catcode `\$12\catcode
  `\&12\catcode `\#12\catcode `\^12\catcode `\_12\catcode `\%12\relax}%
\providecommand \@@startlink[1]{}%
\providecommand \@@endlink[0]{}%
\providecommand \url  [0]{\begingroup\@sanitize@url \@url }%
\providecommand \@url [1]{\endgroup\@href {#1}{\urlprefix }}%
\providecommand \urlprefix  [0]{URL }%
\providecommand \Eprint [0]{\href }%
\providecommand \doibase [0]{http://dx.doi.org/}%
\providecommand \selectlanguage [0]{\@gobble}%
\providecommand \bibinfo  [0]{\@secondoftwo}%
\providecommand \bibfield  [0]{\@secondoftwo}%
\providecommand \translation [1]{[#1]}%
\providecommand \BibitemOpen [0]{}%
\providecommand \bibitemStop [0]{}%
\providecommand \bibitemNoStop [0]{.\EOS\space}%
\providecommand \EOS [0]{\spacefactor3000\relax}%
\providecommand \BibitemShut  [1]{\csname bibitem#1\endcsname}%
\let\auto@bib@innerbib\@empty
\bibitem [{\citenamefont {Coulter}\ \emph {et~al.}(2017)\citenamefont {Coulter}
  \emph {et~al.}}]{Coulter:2017wya}%
  \BibitemOpen
  \bibfield  {author} {\bibinfo {author} {\bibfnamefont {D.~A.}\ \bibnamefont
  {Coulter}} \emph {et~al.},\ }\href {\doibase 10.1126/science.aap9811}
  {\bibfield  {journal} {\bibinfo  {journal} {Science}\ } (\bibinfo {year}
  {2017}),\ 10.1126/science.aap9811},\ \bibinfo {note}
  {[Science358,1556(2017)]},\ \Eprint {http://arxiv.org/abs/1710.05452}
  {arXiv:1710.05452 [astro-ph.HE]} \BibitemShut {NoStop}%
\bibitem [{\citenamefont {Chornock}\ \emph {et~al.}(2017)\citenamefont
  {Chornock} \emph {et~al.}}]{Chornock:2017sdf}%
  \BibitemOpen
  \bibfield  {author} {\bibinfo {author} {\bibfnamefont {R.}~\bibnamefont
  {Chornock}} \emph {et~al.},\ }\href {\doibase 10.3847/2041-8213/aa905c}
  {\bibfield  {journal} {\bibinfo  {journal} {Astrophys. J.}\ }\textbf
  {\bibinfo {volume} {848}},\ \bibinfo {pages} {L19} (\bibinfo {year}
  {2017})},\ \Eprint {http://arxiv.org/abs/1710.05454} {arXiv:1710.05454
  [astro-ph.HE]} \BibitemShut {NoStop}%
\bibitem [{\citenamefont {Nicholl}\ \emph {et~al.}(2017)\citenamefont {Nicholl}
  \emph {et~al.}}]{Nicholl:2017ahq}%
  \BibitemOpen
  \bibfield  {author} {\bibinfo {author} {\bibfnamefont {M.}~\bibnamefont
  {Nicholl}} \emph {et~al.},\ }\href {\doibase 10.3847/2041-8213/aa9029}
  {\bibfield  {journal} {\bibinfo  {journal} {Astrophys. J.}\ }\textbf
  {\bibinfo {volume} {848}},\ \bibinfo {pages} {L18} (\bibinfo {year}
  {2017})},\ \Eprint {http://arxiv.org/abs/1710.05456} {arXiv:1710.05456
  [astro-ph.HE]} \BibitemShut {NoStop}%
\bibitem [{\citenamefont {Cowperthwaite}\ \emph {et~al.}(2017)\citenamefont
  {Cowperthwaite} \emph {et~al.}}]{Cowperthwaite:2017dyu}%
  \BibitemOpen
  \bibfield  {author} {\bibinfo {author} {\bibfnamefont {P.~S.}\ \bibnamefont
  {Cowperthwaite}} \emph {et~al.},\ }\href {\doibase 10.3847/2041-8213/aa8fc7}
  {\bibfield  {journal} {\bibinfo  {journal} {Astrophys. J.}\ }\textbf
  {\bibinfo {volume} {848}},\ \bibinfo {pages} {L17} (\bibinfo {year}
  {2017})},\ \Eprint {http://arxiv.org/abs/1710.05840} {arXiv:1710.05840
  [astro-ph.HE]} \BibitemShut {NoStop}%
\bibitem [{\citenamefont {Tanvir}\ \emph {et~al.}(2017)\citenamefont {Tanvir}
  \emph {et~al.}}]{Tanvir:2017pws}%
  \BibitemOpen
  \bibfield  {author} {\bibinfo {author} {\bibfnamefont {N.~R.}\ \bibnamefont
  {Tanvir}} \emph {et~al.},\ }\href {\doibase 10.3847/2041-8213/aa90b6}
  {\bibfield  {journal} {\bibinfo  {journal} {Astrophys. J.}\ }\textbf
  {\bibinfo {volume} {848}},\ \bibinfo {pages} {L27} (\bibinfo {year}
  {2017})},\ \Eprint {http://arxiv.org/abs/1710.05455} {arXiv:1710.05455
  [astro-ph.HE]} \BibitemShut {NoStop}%
\bibitem [{\citenamefont {Tanaka}\ \emph {et~al.}(2017)\citenamefont {Tanaka}
  \emph {et~al.}}]{Tanaka:2017qxj}%
  \BibitemOpen
  \bibfield  {author} {\bibinfo {author} {\bibfnamefont {M.}~\bibnamefont
  {Tanaka}} \emph {et~al.},\ }\href {\doibase 10.1093/pasj/psx121} {\bibfield
  {journal} {\bibinfo  {journal} {Publ. Astron. Soc. Jap.}\ } (\bibinfo {year}
  {2017}),\ 10.1093/pasj/psx121},\ \Eprint {http://arxiv.org/abs/1710.05850}
  {arXiv:1710.05850 [astro-ph.HE]} \BibitemShut {NoStop}%
\bibitem [{\citenamefont {Abbott}\ \emph {et~al.}(2017)\citenamefont {Abbott}
  \emph {et~al.}}]{TheLIGOScientific:2017qsa}%
  \BibitemOpen
  \bibfield  {author} {\bibinfo {author} {\bibfnamefont {B.~P.}\ \bibnamefont
  {Abbott}} \emph {et~al.} (\bibinfo {collaboration} {Virgo, LIGO
  Scientific}),\ }\href {\doibase 10.1103/PhysRevLett.119.161101} {\bibfield
  {journal} {\bibinfo  {journal} {Phys. Rev. Lett.}\ }\textbf {\bibinfo
  {volume} {119}},\ \bibinfo {pages} {161101} (\bibinfo {year} {2017})},\
  \Eprint {http://arxiv.org/abs/1710.05832} {arXiv:1710.05832 [gr-qc]}
  \BibitemShut {NoStop}%
\bibitem [{\citenamefont {{Lattimer}}\ and\ \citenamefont
  {{Schramm}}(1974)}]{Lattimer:1974a}%
  \BibitemOpen
  \bibfield  {author} {\bibinfo {author} {\bibfnamefont {J.~M.}\ \bibnamefont
  {{Lattimer}}}\ and\ \bibinfo {author} {\bibfnamefont {D.~N.}\ \bibnamefont
  {{Schramm}}},\ }\href {\doibase 10.1086/181612} {\bibfield  {journal}
  {\bibinfo  {journal} {apjl}\ }\textbf {\bibinfo {volume} {192}},\ \bibinfo
  {pages} {L145} (\bibinfo {year} {1974})}\BibitemShut {NoStop}%
\bibitem [{\citenamefont {Li}\ and\ \citenamefont
  {Paczynski}(1998)}]{Li:1998bw}%
  \BibitemOpen
  \bibfield  {author} {\bibinfo {author} {\bibfnamefont {L.-X.}\ \bibnamefont
  {Li}}\ and\ \bibinfo {author} {\bibfnamefont {B.}~\bibnamefont {Paczynski}},\
  }\href {\doibase 10.1086/311680} {\bibfield  {journal} {\bibinfo  {journal}
  {Astrophys.J.}\ }\textbf {\bibinfo {volume} {507}},\ \bibinfo {pages} {L59}
  (\bibinfo {year} {1998})},\ \Eprint {http://arxiv.org/abs/astro-ph/9807272}
  {arXiv:astro-ph/9807272 [astro-ph]} \BibitemShut {NoStop}%
\bibitem [{\citenamefont {Kulkarni}(2005)}]{Kulkarni:2005jw}%
  \BibitemOpen
  \bibfield  {author} {\bibinfo {author} {\bibfnamefont {S.~R.}\ \bibnamefont
  {Kulkarni}},\ }\href@noop {} {\  (\bibinfo {year} {2005})},\ \Eprint
  {http://arxiv.org/abs/astro-ph/0510256} {arXiv:astro-ph/0510256 [astro-ph]}
  \BibitemShut {NoStop}%
\bibitem [{\citenamefont {Rosswog}(2005)}]{Rosswog:2005su}%
  \BibitemOpen
  \bibfield  {author} {\bibinfo {author} {\bibfnamefont {S.}~\bibnamefont
  {Rosswog}},\ }\href {\doibase 10.1086/497062} {\bibfield  {journal} {\bibinfo
   {journal} {Astrophys. J.}\ }\textbf {\bibinfo {volume} {634}},\ \bibinfo
  {pages} {1202} (\bibinfo {year} {2005})},\ \Eprint
  {http://arxiv.org/abs/astro-ph/0508138} {arXiv:astro-ph/0508138 [astro-ph]}
  \BibitemShut {NoStop}%
\bibitem [{\citenamefont {Metzger}\ \emph {et~al.}(2010)\citenamefont
  {Metzger}, \citenamefont {Martinez-Pinedo}, \citenamefont {Darbha},
  \citenamefont {Quataert}, \citenamefont {Arcones} \emph
  {et~al.}}]{Metzger:2010sy}%
  \BibitemOpen
  \bibfield  {author} {\bibinfo {author} {\bibfnamefont {B.}~\bibnamefont
  {Metzger}}, \bibinfo {author} {\bibfnamefont {G.}~\bibnamefont
  {Martinez-Pinedo}}, \bibinfo {author} {\bibfnamefont {S.}~\bibnamefont
  {Darbha}}, \bibinfo {author} {\bibfnamefont {E.}~\bibnamefont {Quataert}},
  \bibinfo {author} {\bibfnamefont {A.}~\bibnamefont {Arcones}},  \emph
  {et~al.},\ }\href {\doibase 10.1111/j.1365-2966.2010.16864.x} {\bibfield
  {journal} {\bibinfo  {journal} {Mon.Not.Roy.Astron.Soc.}\ }\textbf {\bibinfo
  {volume} {406}},\ \bibinfo {pages} {2650} (\bibinfo {year} {2010})},\ \Eprint
  {http://arxiv.org/abs/1001.5029} {arXiv:1001.5029 [astro-ph.HE]} \BibitemShut
  {NoStop}%
\bibitem [{\citenamefont {Roberts}\ \emph {et~al.}(2011)\citenamefont
  {Roberts}, \citenamefont {Kasen}, \citenamefont {Lee},\ and\ \citenamefont
  {Ramirez-Ruiz}}]{Roberts:2011xz}%
  \BibitemOpen
  \bibfield  {author} {\bibinfo {author} {\bibfnamefont {L.~F.}\ \bibnamefont
  {Roberts}}, \bibinfo {author} {\bibfnamefont {D.}~\bibnamefont {Kasen}},
  \bibinfo {author} {\bibfnamefont {W.~H.}\ \bibnamefont {Lee}}, \ and\
  \bibinfo {author} {\bibfnamefont {E.}~\bibnamefont {Ramirez-Ruiz}},\ }\href
  {\doibase 10.1088/2041-8205/736/1/L21} {\bibfield  {journal} {\bibinfo
  {journal} {Astrophys.J.}\ }\textbf {\bibinfo {volume} {736}},\ \bibinfo
  {pages} {L21} (\bibinfo {year} {2011})},\ \Eprint
  {http://arxiv.org/abs/1104.5504} {arXiv:1104.5504 [astro-ph.HE]} \BibitemShut
  {NoStop}%
\bibitem [{\citenamefont {Kasen}\ \emph {et~al.}(2013)\citenamefont {Kasen},
  \citenamefont {Badnell},\ and\ \citenamefont {Barnes}}]{Kasen:2013xka}%
  \BibitemOpen
  \bibfield  {author} {\bibinfo {author} {\bibfnamefont {D.}~\bibnamefont
  {Kasen}}, \bibinfo {author} {\bibfnamefont {N.~R.}\ \bibnamefont {Badnell}},
  \ and\ \bibinfo {author} {\bibfnamefont {J.}~\bibnamefont {Barnes}},\ }\href
  {\doibase 10.1088/0004-637X/774/1/25} {\bibfield  {journal} {\bibinfo
  {journal} {Astrophys. J.}\ }\textbf {\bibinfo {volume} {774}},\ \bibinfo
  {pages} {25} (\bibinfo {year} {2013})},\ \Eprint
  {http://arxiv.org/abs/1303.5788} {arXiv:1303.5788 [astro-ph.HE]} \BibitemShut
  {NoStop}%
\bibitem [{\citenamefont {Martin}\ \emph {et~al.}(2015)\citenamefont {Martin},
  \citenamefont {Perego}, \citenamefont {Arcones}, \citenamefont {Thielemann},
  \citenamefont {Korobkin},\ and\ \citenamefont {Rosswog}}]{Martin:2015hxa}%
  \BibitemOpen
  \bibfield  {author} {\bibinfo {author} {\bibfnamefont {D.}~\bibnamefont
  {Martin}}, \bibinfo {author} {\bibfnamefont {A.}~\bibnamefont {Perego}},
  \bibinfo {author} {\bibfnamefont {A.}~\bibnamefont {Arcones}}, \bibinfo
  {author} {\bibfnamefont {F.-K.}\ \bibnamefont {Thielemann}}, \bibinfo
  {author} {\bibfnamefont {O.}~\bibnamefont {Korobkin}}, \ and\ \bibinfo
  {author} {\bibfnamefont {S.}~\bibnamefont {Rosswog}},\ }\href {\doibase
  10.1088/0004-637X/813/1/2} {\bibfield  {journal} {\bibinfo  {journal}
  {Astrophys. J.}\ }\textbf {\bibinfo {volume} {813}},\ \bibinfo {pages} {2}
  (\bibinfo {year} {2015})},\ \Eprint {http://arxiv.org/abs/1506.05048}
  {arXiv:1506.05048 [astro-ph.SR]} \BibitemShut {NoStop}%
\bibitem [{\citenamefont {Villar}\ \emph {et~al.}(2017)\citenamefont {Villar}
  \emph {et~al.}}]{Villar:2017wcc}%
  \BibitemOpen
  \bibfield  {author} {\bibinfo {author} {\bibfnamefont {V.~A.}\ \bibnamefont
  {Villar}} \emph {et~al.},\ }\href {\doibase 10.3847/2041-8213/aa9c84}
  {\bibfield  {journal} {\bibinfo  {journal} {Astrophys. J.}\ }\textbf
  {\bibinfo {volume} {851}},\ \bibinfo {pages} {L21} (\bibinfo {year}
  {2017})},\ \Eprint {http://arxiv.org/abs/1710.11576} {arXiv:1710.11576
  [astro-ph.HE]} \BibitemShut {NoStop}%
\bibitem [{\citenamefont {Waxman}\ \emph {et~al.}(2018)\citenamefont {Waxman},
  \citenamefont {Ofek}, \citenamefont {Kushnir},\ and\ \citenamefont
  {Gal-Yam}}]{Waxman:2017sqv}%
  \BibitemOpen
  \bibfield  {author} {\bibinfo {author} {\bibfnamefont {E.}~\bibnamefont
  {Waxman}}, \bibinfo {author} {\bibfnamefont {E.~O.}\ \bibnamefont {Ofek}},
  \bibinfo {author} {\bibfnamefont {D.}~\bibnamefont {Kushnir}}, \ and\
  \bibinfo {author} {\bibfnamefont {A.}~\bibnamefont {Gal-Yam}},\ }\href
  {\doibase 10.1093/mnras/sty2441} {\bibfield  {journal} {\bibinfo  {journal}
  {Mon. Not. Roy. Astron. Soc.}\ }\textbf {\bibinfo {volume} {481}},\ \bibinfo
  {pages} {3423} (\bibinfo {year} {2018})},\ \Eprint
  {http://arxiv.org/abs/1711.09638} {arXiv:1711.09638 [astro-ph.HE]}
  \BibitemShut {NoStop}%
\bibitem [{\citenamefont {Hotokezaka}\ \emph {et~al.}(2013)\citenamefont
  {Hotokezaka}, \citenamefont {Kiuchi}, \citenamefont {Kyutoku}, \citenamefont
  {Okawa}, \citenamefont {Sekiguchi} \emph {et~al.}}]{Hotokezaka:2012ze}%
  \BibitemOpen
  \bibfield  {author} {\bibinfo {author} {\bibfnamefont {K.}~\bibnamefont
  {Hotokezaka}}, \bibinfo {author} {\bibfnamefont {K.}~\bibnamefont {Kiuchi}},
  \bibinfo {author} {\bibfnamefont {K.}~\bibnamefont {Kyutoku}}, \bibinfo
  {author} {\bibfnamefont {H.}~\bibnamefont {Okawa}}, \bibinfo {author}
  {\bibfnamefont {Y.-i.}\ \bibnamefont {Sekiguchi}},  \emph {et~al.},\ }\href
  {\doibase 10.1103/PhysRevD.87.024001} {\bibfield  {journal} {\bibinfo
  {journal} {Phys.Rev.}\ }\textbf {\bibinfo {volume} {D87}},\ \bibinfo {pages}
  {024001} (\bibinfo {year} {2013})},\ \Eprint {http://arxiv.org/abs/1212.0905}
  {arXiv:1212.0905 [astro-ph.HE]} \BibitemShut {NoStop}%
\bibitem [{\citenamefont {Bauswein}\ \emph {et~al.}(2013)\citenamefont
  {Bauswein}, \citenamefont {Goriely},\ and\ \citenamefont
  {Janka}}]{Bauswein:2013yna}%
  \BibitemOpen
  \bibfield  {author} {\bibinfo {author} {\bibfnamefont {A.}~\bibnamefont
  {Bauswein}}, \bibinfo {author} {\bibfnamefont {S.}~\bibnamefont {Goriely}}, \
  and\ \bibinfo {author} {\bibfnamefont {H.-T.}\ \bibnamefont {Janka}},\ }\href
  {\doibase 10.1088/0004-637X/773/1/78} {\bibfield  {journal} {\bibinfo
  {journal} {Astrophys.J.}\ }\textbf {\bibinfo {volume} {773}},\ \bibinfo
  {pages} {78} (\bibinfo {year} {2013})},\ \Eprint
  {http://arxiv.org/abs/1302.6530} {arXiv:1302.6530 [astro-ph.SR]} \BibitemShut
  {NoStop}%
\bibitem [{\citenamefont {Radice}\ \emph
  {et~al.}(2018{\natexlab{a}})\citenamefont {Radice}, \citenamefont {Perego},
  \citenamefont {Hotokezaka}, \citenamefont {Fromm}, \citenamefont {Bernuzzi},\
  and\ \citenamefont {Roberts}}]{Radice:2018pdn}%
  \BibitemOpen
  \bibfield  {author} {\bibinfo {author} {\bibfnamefont {D.}~\bibnamefont
  {Radice}}, \bibinfo {author} {\bibfnamefont {A.}~\bibnamefont {Perego}},
  \bibinfo {author} {\bibfnamefont {K.}~\bibnamefont {Hotokezaka}}, \bibinfo
  {author} {\bibfnamefont {S.~A.}\ \bibnamefont {Fromm}}, \bibinfo {author}
  {\bibfnamefont {S.}~\bibnamefont {Bernuzzi}}, \ and\ \bibinfo {author}
  {\bibfnamefont {L.~F.}\ \bibnamefont {Roberts}},\ }\href {\doibase
  10.3847/1538-4357/aaf054} {\bibfield  {journal} {\bibinfo  {journal}
  {Astrophys. J.}\ }\textbf {\bibinfo {volume} {869}},\ \bibinfo {pages} {130}
  (\bibinfo {year} {2018}{\natexlab{a}})},\ \Eprint
  {http://arxiv.org/abs/1809.11161} {arXiv:1809.11161 [astro-ph.HE]}
  \BibitemShut {NoStop}%
\bibitem [{\citenamefont {Sekiguchi}\ \emph {et~al.}(2015)\citenamefont
  {Sekiguchi}, \citenamefont {Kiuchi}, \citenamefont {Kyutoku},\ and\
  \citenamefont {Shibata}}]{Sekiguchi:2015dma}%
  \BibitemOpen
  \bibfield  {author} {\bibinfo {author} {\bibfnamefont {Y.}~\bibnamefont
  {Sekiguchi}}, \bibinfo {author} {\bibfnamefont {K.}~\bibnamefont {Kiuchi}},
  \bibinfo {author} {\bibfnamefont {K.}~\bibnamefont {Kyutoku}}, \ and\
  \bibinfo {author} {\bibfnamefont {M.}~\bibnamefont {Shibata}},\ }\href
  {\doibase 10.1103/PhysRevD.91.064059} {\bibfield  {journal} {\bibinfo
  {journal} {Phys.Rev.}\ }\textbf {\bibinfo {volume} {D91}},\ \bibinfo {pages}
  {064059} (\bibinfo {year} {2015})},\ \Eprint
  {http://arxiv.org/abs/1502.06660} {arXiv:1502.06660 [astro-ph.HE]}
  \BibitemShut {NoStop}%
\bibitem [{\citenamefont {Perego}\ \emph {et~al.}(2014)\citenamefont {Perego},
  \citenamefont {Rosswog}, \citenamefont {Cabezon}, \citenamefont {Korobkin},
  \citenamefont {Kaeppeli} \emph {et~al.}}]{Perego:2014fma}%
  \BibitemOpen
  \bibfield  {author} {\bibinfo {author} {\bibfnamefont {A.}~\bibnamefont
  {Perego}}, \bibinfo {author} {\bibfnamefont {S.}~\bibnamefont {Rosswog}},
  \bibinfo {author} {\bibfnamefont {R.}~\bibnamefont {Cabezon}}, \bibinfo
  {author} {\bibfnamefont {O.}~\bibnamefont {Korobkin}}, \bibinfo {author}
  {\bibfnamefont {R.}~\bibnamefont {Kaeppeli}},  \emph {et~al.},\ }\href
  {\doibase 10.1093/mnras/stu1352} {\bibfield  {journal} {\bibinfo  {journal}
  {Mon.Not.Roy.Astron.Soc.}\ }\textbf {\bibinfo {volume} {443}},\ \bibinfo
  {pages} {3134} (\bibinfo {year} {2014})},\ \Eprint
  {http://arxiv.org/abs/1405.6730} {arXiv:1405.6730 [astro-ph.HE]} \BibitemShut
  {NoStop}%
\bibitem [{\citenamefont {Just}\ \emph {et~al.}(2015)\citenamefont {Just},
  \citenamefont {Bauswein}, \citenamefont {Pulpillo}, \citenamefont {Goriely},\
  and\ \citenamefont {Janka}}]{Just:2014fka}%
  \BibitemOpen
  \bibfield  {author} {\bibinfo {author} {\bibfnamefont {O.}~\bibnamefont
  {Just}}, \bibinfo {author} {\bibfnamefont {A.}~\bibnamefont {Bauswein}},
  \bibinfo {author} {\bibfnamefont {R.~A.}\ \bibnamefont {Pulpillo}}, \bibinfo
  {author} {\bibfnamefont {S.}~\bibnamefont {Goriely}}, \ and\ \bibinfo
  {author} {\bibfnamefont {H.~T.}\ \bibnamefont {Janka}},\ }\href {\doibase
  10.1093/mnras/stv009} {\bibfield  {journal} {\bibinfo  {journal} {Mon. Not.
  Roy. Astron. Soc.}\ }\textbf {\bibinfo {volume} {448}},\ \bibinfo {pages}
  {541} (\bibinfo {year} {2015})},\ \Eprint {http://arxiv.org/abs/1406.2687}
  {arXiv:1406.2687 [astro-ph.SR]} \BibitemShut {NoStop}%
\bibitem [{\citenamefont {Kasen}\ \emph {et~al.}(2015)\citenamefont {Kasen},
  \citenamefont {Fern{\'a}ndez},\ and\ \citenamefont
  {Metzger}}]{Kasen:2014toa}%
  \BibitemOpen
  \bibfield  {author} {\bibinfo {author} {\bibfnamefont {D.}~\bibnamefont
  {Kasen}}, \bibinfo {author} {\bibfnamefont {R.}~\bibnamefont
  {Fern{\'a}ndez}}, \ and\ \bibinfo {author} {\bibfnamefont {B.}~\bibnamefont
  {Metzger}},\ }\href {\doibase 10.1093/mnras/stv721} {\bibfield  {journal}
  {\bibinfo  {journal} {Mon. Not. Roy. Astron. Soc.}\ }\textbf {\bibinfo
  {volume} {450}},\ \bibinfo {pages} {1777} (\bibinfo {year} {2015})},\ \Eprint
  {http://arxiv.org/abs/1411.3726} {arXiv:1411.3726 [astro-ph.HE]} \BibitemShut
  {NoStop}%
\bibitem [{\citenamefont {Metzger}\ and\ \citenamefont
  {Fern\'{a}ndez}(2014)}]{Metzger:2014ila}%
  \BibitemOpen
  \bibfield  {author} {\bibinfo {author} {\bibfnamefont {B.~D.}\ \bibnamefont
  {Metzger}}\ and\ \bibinfo {author} {\bibfnamefont {R.}~\bibnamefont
  {Fern\'{a}ndez}},\ }\href {\doibase 10.1093/mnras/stu802} {\bibfield
  {journal} {\bibinfo  {journal} {Mon.Not.Roy.Astron.Soc.}\ }\textbf {\bibinfo
  {volume} {441}},\ \bibinfo {pages} {3444} (\bibinfo {year} {2014})},\ \Eprint
  {http://arxiv.org/abs/1402.4803} {arXiv:1402.4803 [astro-ph.HE]} \BibitemShut
  {NoStop}%
\bibitem [{\citenamefont {Wu}\ \emph {et~al.}(2016)\citenamefont {Wu},
  \citenamefont {Fern\'{a}ndez}, \citenamefont {Martínez-Pinedo},\ and\
  \citenamefont {Metzger}}]{Wu:2016pnw}%
  \BibitemOpen
  \bibfield  {author} {\bibinfo {author} {\bibfnamefont {M.-R.}\ \bibnamefont
  {Wu}}, \bibinfo {author} {\bibfnamefont {R.}~\bibnamefont {Fern\'{a}ndez}},
  \bibinfo {author} {\bibfnamefont {G.}~\bibnamefont {Martínez-Pinedo}}, \
  and\ \bibinfo {author} {\bibfnamefont {B.~D.}\ \bibnamefont {Metzger}},\
  }\href {\doibase 10.1093/mnras/stw2156} {\bibfield  {journal} {\bibinfo
  {journal} {Mon. Not. Roy. Astron. Soc.}\ }\textbf {\bibinfo {volume} {463}},\
  \bibinfo {pages} {2323} (\bibinfo {year} {2016})},\ \Eprint
  {http://arxiv.org/abs/1607.05290} {arXiv:1607.05290 [astro-ph.HE]}
  \BibitemShut {NoStop}%
\bibitem [{\citenamefont {Siegel}\ and\ \citenamefont
  {Metzger}(2017)}]{Siegel:2017nub}%
  \BibitemOpen
  \bibfield  {author} {\bibinfo {author} {\bibfnamefont {D.~M.}\ \bibnamefont
  {Siegel}}\ and\ \bibinfo {author} {\bibfnamefont {B.~D.}\ \bibnamefont
  {Metzger}},\ }\href {\doibase 10.1103/PhysRevLett.119.231102} {\bibfield
  {journal} {\bibinfo  {journal} {Phys. Rev. Lett.}\ }\textbf {\bibinfo
  {volume} {119}},\ \bibinfo {pages} {231102} (\bibinfo {year} {2017})},\
  \Eprint {http://arxiv.org/abs/1705.05473} {arXiv:1705.05473 [astro-ph.HE]}
  \BibitemShut {NoStop}%
\bibitem [{\citenamefont {Fujibayashi}\ \emph {et~al.}(2018)\citenamefont
  {Fujibayashi}, \citenamefont {Kiuchi}, \citenamefont {Nishimura},
  \citenamefont {Sekiguchi},\ and\ \citenamefont
  {Shibata}}]{Fujibayashi:2017puw}%
  \BibitemOpen
  \bibfield  {author} {\bibinfo {author} {\bibfnamefont {S.}~\bibnamefont
  {Fujibayashi}}, \bibinfo {author} {\bibfnamefont {K.}~\bibnamefont {Kiuchi}},
  \bibinfo {author} {\bibfnamefont {N.}~\bibnamefont {Nishimura}}, \bibinfo
  {author} {\bibfnamefont {Y.}~\bibnamefont {Sekiguchi}}, \ and\ \bibinfo
  {author} {\bibfnamefont {M.}~\bibnamefont {Shibata}},\ }\href {\doibase
  10.3847/1538-4357/aabafd} {\bibfield  {journal} {\bibinfo  {journal}
  {Astrophys. J.}\ }\textbf {\bibinfo {volume} {860}},\ \bibinfo {pages} {64}
  (\bibinfo {year} {2018})},\ \Eprint {http://arxiv.org/abs/1711.02093}
  {arXiv:1711.02093 [astro-ph.HE]} \BibitemShut {NoStop}%
\bibitem [{\citenamefont {Miller}\ \emph {et~al.}(2019)\citenamefont {Miller},
  \citenamefont {Ryan}, \citenamefont {Dolence}, \citenamefont {Burrows},
  \citenamefont {Fontes}, \citenamefont {Fryer}, \citenamefont {Korobkin},
  \citenamefont {Lippuner}, \citenamefont {Mumpower},\ and\ \citenamefont
  {Wollaeger}}]{Miller:2019dpt}%
  \BibitemOpen
  \bibfield  {author} {\bibinfo {author} {\bibfnamefont {J.~M.}\ \bibnamefont
  {Miller}}, \bibinfo {author} {\bibfnamefont {B.~R.}\ \bibnamefont {Ryan}},
  \bibinfo {author} {\bibfnamefont {J.~C.}\ \bibnamefont {Dolence}}, \bibinfo
  {author} {\bibfnamefont {A.}~\bibnamefont {Burrows}}, \bibinfo {author}
  {\bibfnamefont {C.~J.}\ \bibnamefont {Fontes}}, \bibinfo {author}
  {\bibfnamefont {C.~L.}\ \bibnamefont {Fryer}}, \bibinfo {author}
  {\bibfnamefont {O.}~\bibnamefont {Korobkin}}, \bibinfo {author}
  {\bibfnamefont {J.}~\bibnamefont {Lippuner}}, \bibinfo {author}
  {\bibfnamefont {M.~R.}\ \bibnamefont {Mumpower}}, \ and\ \bibinfo {author}
  {\bibfnamefont {R.~T.}\ \bibnamefont {Wollaeger}},\ }\href {\doibase
  10.1103/PhysRevD.100.023008} {\bibfield  {journal} {\bibinfo  {journal}
  {Phys. Rev.}\ }\textbf {\bibinfo {volume} {D100}},\ \bibinfo {pages} {023008}
  (\bibinfo {year} {2019})},\ \Eprint {http://arxiv.org/abs/1905.07477}
  {arXiv:1905.07477 [astro-ph.HE]} \BibitemShut {NoStop}%
\bibitem [{\citenamefont {Radice}\ \emph
  {et~al.}(2018{\natexlab{b}})\citenamefont {Radice}, \citenamefont {Perego},
  \citenamefont {Zappa},\ and\ \citenamefont {Bernuzzi}}]{Radice:2017lry}%
  \BibitemOpen
  \bibfield  {author} {\bibinfo {author} {\bibfnamefont {D.}~\bibnamefont
  {Radice}}, \bibinfo {author} {\bibfnamefont {A.}~\bibnamefont {Perego}},
  \bibinfo {author} {\bibfnamefont {F.}~\bibnamefont {Zappa}}, \ and\ \bibinfo
  {author} {\bibfnamefont {S.}~\bibnamefont {Bernuzzi}},\ }\href {\doibase
  10.3847/2041-8213/aaa402} {\bibfield  {journal} {\bibinfo  {journal}
  {Astrophys. J.}\ }\textbf {\bibinfo {volume} {852}},\ \bibinfo {pages} {L29}
  (\bibinfo {year} {2018}{\natexlab{b}})},\ \Eprint
  {http://arxiv.org/abs/1711.03647} {arXiv:1711.03647 [astro-ph.HE]}
  \BibitemShut {NoStop}%
\bibitem [{\citenamefont {Perego}\ \emph {et~al.}(2019)\citenamefont {Perego},
  \citenamefont {Bernuzzi},\ and\ \citenamefont {Radice}}]{Perego:2019adq}%
  \BibitemOpen
  \bibfield  {author} {\bibinfo {author} {\bibfnamefont {A.}~\bibnamefont
  {Perego}}, \bibinfo {author} {\bibfnamefont {S.}~\bibnamefont {Bernuzzi}}, \
  and\ \bibinfo {author} {\bibfnamefont {D.}~\bibnamefont {Radice}},\ }\href
  {\doibase 10.1140/epja/i2019-12810-7} {\bibfield  {journal} {\bibinfo
  {journal} {Eur. Phys. J.}\ }\textbf {\bibinfo {volume} {A55}},\ \bibinfo
  {pages} {124} (\bibinfo {year} {2019})},\ \Eprint
  {http://arxiv.org/abs/1903.07898} {arXiv:1903.07898 [gr-qc]} \BibitemShut
  {NoStop}%
\bibitem [{\citenamefont {Dessart}\ \emph {et~al.}(2009)\citenamefont
  {Dessart}, \citenamefont {Ott}, \citenamefont {Burrows}, \citenamefont
  {Rosswog},\ and\ \citenamefont {Livne}}]{Dessart:2008zd}%
  \BibitemOpen
  \bibfield  {author} {\bibinfo {author} {\bibfnamefont {L.}~\bibnamefont
  {Dessart}}, \bibinfo {author} {\bibfnamefont {C.}~\bibnamefont {Ott}},
  \bibinfo {author} {\bibfnamefont {A.}~\bibnamefont {Burrows}}, \bibinfo
  {author} {\bibfnamefont {S.}~\bibnamefont {Rosswog}}, \ and\ \bibinfo
  {author} {\bibfnamefont {E.}~\bibnamefont {Livne}},\ }\href {\doibase
  10.1088/0004-637X/690/2/1681} {\bibfield  {journal} {\bibinfo  {journal}
  {Astrophys.J.}\ }\textbf {\bibinfo {volume} {690}},\ \bibinfo {pages} {1681}
  (\bibinfo {year} {2009})},\ \Eprint {http://arxiv.org/abs/0806.4380}
  {arXiv:0806.4380 [astro-ph]} \BibitemShut {NoStop}%
\bibitem [{\citenamefont {Fern\'{a}ndez}\ \emph {et~al.}(2015)\citenamefont
  {Fern\'{a}ndez}, \citenamefont {Quataert}, \citenamefont {Schwab},
  \citenamefont {Kasen},\ and\ \citenamefont {Rosswog}}]{Fernandez:2014bra}%
  \BibitemOpen
  \bibfield  {author} {\bibinfo {author} {\bibfnamefont {R.}~\bibnamefont
  {Fern\'{a}ndez}}, \bibinfo {author} {\bibfnamefont {E.}~\bibnamefont
  {Quataert}}, \bibinfo {author} {\bibfnamefont {J.}~\bibnamefont {Schwab}},
  \bibinfo {author} {\bibfnamefont {D.}~\bibnamefont {Kasen}}, \ and\ \bibinfo
  {author} {\bibfnamefont {S.}~\bibnamefont {Rosswog}},\ }\href {\doibase
  10.1093/mnras/stv238} {\bibfield  {journal} {\bibinfo  {journal} {Mon. Not.
  Roy. Astron. Soc.}\ }\textbf {\bibinfo {volume} {449}},\ \bibinfo {pages}
  {390} (\bibinfo {year} {2015})},\ \Eprint {http://arxiv.org/abs/1412.5588}
  {arXiv:1412.5588 [astro-ph.HE]} \BibitemShut {NoStop}%
\bibitem [{\citenamefont {Lippuner}\ \emph {et~al.}(2017)\citenamefont
  {Lippuner}, \citenamefont {Fern\'{a}ndez}, \citenamefont {Roberts},
  \citenamefont {Foucart}, \citenamefont {Kasen}, \citenamefont {Metzger},\
  and\ \citenamefont {Ott}}]{Lippuner:2017bfm}%
  \BibitemOpen
  \bibfield  {author} {\bibinfo {author} {\bibfnamefont {J.}~\bibnamefont
  {Lippuner}}, \bibinfo {author} {\bibfnamefont {R.}~\bibnamefont
  {Fern\'{a}ndez}}, \bibinfo {author} {\bibfnamefont {L.~F.}\ \bibnamefont
  {Roberts}}, \bibinfo {author} {\bibfnamefont {F.}~\bibnamefont {Foucart}},
  \bibinfo {author} {\bibfnamefont {D.}~\bibnamefont {Kasen}}, \bibinfo
  {author} {\bibfnamefont {B.~D.}\ \bibnamefont {Metzger}}, \ and\ \bibinfo
  {author} {\bibfnamefont {C.~D.}\ \bibnamefont {Ott}},\ }\href {\doibase
  10.1093/mnras/stx1987} {\bibfield  {journal} {\bibinfo  {journal} {Mon. Not.
  Roy. Astron. Soc.}\ }\textbf {\bibinfo {volume} {472}},\ \bibinfo {pages}
  {904} (\bibinfo {year} {2017})},\ \Eprint {http://arxiv.org/abs/1703.06216}
  {arXiv:1703.06216 [astro-ph.HE]} \BibitemShut {NoStop}%
\bibitem [{\citenamefont {Radice}\ \emph
  {et~al.}(2018{\natexlab{c}})\citenamefont {Radice}, \citenamefont {Perego},
  \citenamefont {Bernuzzi},\ and\ \citenamefont {Zhang}}]{Radice:2018xqa}%
  \BibitemOpen
  \bibfield  {author} {\bibinfo {author} {\bibfnamefont {D.}~\bibnamefont
  {Radice}}, \bibinfo {author} {\bibfnamefont {A.}~\bibnamefont {Perego}},
  \bibinfo {author} {\bibfnamefont {S.}~\bibnamefont {Bernuzzi}}, \ and\
  \bibinfo {author} {\bibfnamefont {B.}~\bibnamefont {Zhang}},\ }\href
  {\doibase 10.1093/mnras/sty2531} {\bibfield  {journal} {\bibinfo  {journal}
  {Mon. Not. Roy. Astron. Soc.}\ }\textbf {\bibinfo {volume} {481}},\ \bibinfo
  {pages} {3670} (\bibinfo {year} {2018}{\natexlab{c}})},\ \Eprint
  {http://arxiv.org/abs/1803.10865} {arXiv:1803.10865 [astro-ph.HE]}
  \BibitemShut {NoStop}%
\bibitem [{\citenamefont {Fern{\'a}ndez}\ \emph {et~al.}(2019)\citenamefont
  {Fern{\'a}ndez}, \citenamefont {Tchekhovskoy}, \citenamefont {Quataert},
  \citenamefont {Foucart},\ and\ \citenamefont {Kasen}}]{Fernandez:2018kax}%
  \BibitemOpen
  \bibfield  {author} {\bibinfo {author} {\bibfnamefont {R.}~\bibnamefont
  {Fern{\'a}ndez}}, \bibinfo {author} {\bibfnamefont {A.}~\bibnamefont
  {Tchekhovskoy}}, \bibinfo {author} {\bibfnamefont {E.}~\bibnamefont
  {Quataert}}, \bibinfo {author} {\bibfnamefont {F.}~\bibnamefont {Foucart}}, \
  and\ \bibinfo {author} {\bibfnamefont {D.}~\bibnamefont {Kasen}},\ }\href
  {\doibase 10.1093/mnras/sty2932} {\bibfield  {journal} {\bibinfo  {journal}
  {Mon. Not. Roy. Astron. Soc.}\ }\textbf {\bibinfo {volume} {482}},\ \bibinfo
  {pages} {3373} (\bibinfo {year} {2019})},\ \Eprint
  {http://arxiv.org/abs/1808.00461} {arXiv:1808.00461 [astro-ph.HE]}
  \BibitemShut {NoStop}%
\bibitem [{\citenamefont {Fahlman}\ and\ \citenamefont
  {Fern{\'a}ndez}(2018)}]{Fahlman:2018llv}%
  \BibitemOpen
  \bibfield  {author} {\bibinfo {author} {\bibfnamefont {S.}~\bibnamefont
  {Fahlman}}\ and\ \bibinfo {author} {\bibfnamefont {R.}~\bibnamefont
  {Fern{\'a}ndez}},\ }\href {\doibase 10.3847/2041-8213/aaf1ab} {\bibfield
  {journal} {\bibinfo  {journal} {Astrophys. J.}\ }\textbf {\bibinfo {volume}
  {869}},\ \bibinfo {pages} {L3} (\bibinfo {year} {2018})},\ \Eprint
  {http://arxiv.org/abs/1811.08906} {arXiv:1811.08906 [astro-ph.HE]}
  \BibitemShut {NoStop}%
\bibitem [{\citenamefont {Perego}\ \emph {et~al.}(2017)\citenamefont {Perego},
  \citenamefont {Radice},\ and\ \citenamefont {Bernuzzi}}]{Perego:2017wtu}%
  \BibitemOpen
  \bibfield  {author} {\bibinfo {author} {\bibfnamefont {A.}~\bibnamefont
  {Perego}}, \bibinfo {author} {\bibfnamefont {D.}~\bibnamefont {Radice}}, \
  and\ \bibinfo {author} {\bibfnamefont {S.}~\bibnamefont {Bernuzzi}},\ }\href
  {\doibase 10.3847/2041-8213/aa9ab9} {\bibfield  {journal} {\bibinfo
  {journal} {Astrophys. J.}\ }\textbf {\bibinfo {volume} {850}},\ \bibinfo
  {pages} {L37} (\bibinfo {year} {2017})},\ \Eprint
  {http://arxiv.org/abs/1711.03982} {arXiv:1711.03982 [astro-ph.HE]}
  \BibitemShut {NoStop}%
\bibitem [{\citenamefont {Kawaguchi}\ \emph {et~al.}(2018)\citenamefont
  {Kawaguchi}, \citenamefont {Shibata},\ and\ \citenamefont
  {Tanaka}}]{Kawaguchi:2018ptg}%
  \BibitemOpen
  \bibfield  {author} {\bibinfo {author} {\bibfnamefont {K.}~\bibnamefont
  {Kawaguchi}}, \bibinfo {author} {\bibfnamefont {M.}~\bibnamefont {Shibata}},
  \ and\ \bibinfo {author} {\bibfnamefont {M.}~\bibnamefont {Tanaka}},\ }\href
  {\doibase 10.3847/2041-8213/aade02} {\bibfield  {journal} {\bibinfo
  {journal} {Astrophys. J.}\ }\textbf {\bibinfo {volume} {865}},\ \bibinfo
  {pages} {L21} (\bibinfo {year} {2018})},\ \Eprint
  {http://arxiv.org/abs/1806.04088} {arXiv:1806.04088 [astro-ph.HE]}
  \BibitemShut {NoStop}%
\bibitem [{\citenamefont {Lazzati}\ \emph {et~al.}(2017)\citenamefont
  {Lazzati}, \citenamefont {Deich}, \citenamefont {Morsony},\ and\
  \citenamefont {Workman}}]{Lazzati:2016yxl}%
  \BibitemOpen
  \bibfield  {author} {\bibinfo {author} {\bibfnamefont {D.}~\bibnamefont
  {Lazzati}}, \bibinfo {author} {\bibfnamefont {A.}~\bibnamefont {Deich}},
  \bibinfo {author} {\bibfnamefont {B.~J.}\ \bibnamefont {Morsony}}, \ and\
  \bibinfo {author} {\bibfnamefont {J.~C.}\ \bibnamefont {Workman}},\ }\href
  {\doibase 10.1093/mnras/stx1683} {\bibfield  {journal} {\bibinfo  {journal}
  {Mon. Not. Roy. Astron. Soc.}\ }\textbf {\bibinfo {volume} {471}},\ \bibinfo
  {pages} {1652} (\bibinfo {year} {2017})},\ \Eprint
  {http://arxiv.org/abs/1610.01157} {arXiv:1610.01157 [astro-ph.HE]}
  \BibitemShut {NoStop}%
\bibitem [{\citenamefont {Bromberg}\ \emph {et~al.}(2018)\citenamefont
  {Bromberg}, \citenamefont {Tchekhovskoy}, \citenamefont {Gottlieb},
  \citenamefont {Nakar},\ and\ \citenamefont {Piran}}]{Bromberg:2017crh}%
  \BibitemOpen
  \bibfield  {author} {\bibinfo {author} {\bibfnamefont {O.}~\bibnamefont
  {Bromberg}}, \bibinfo {author} {\bibfnamefont {A.}~\bibnamefont
  {Tchekhovskoy}}, \bibinfo {author} {\bibfnamefont {O.}~\bibnamefont
  {Gottlieb}}, \bibinfo {author} {\bibfnamefont {E.}~\bibnamefont {Nakar}}, \
  and\ \bibinfo {author} {\bibfnamefont {T.}~\bibnamefont {Piran}},\ }\href
  {\doibase 10.1093/mnras/stx3316} {\bibfield  {journal} {\bibinfo  {journal}
  {Mon. Not. Roy. Astron. Soc.}\ }\textbf {\bibinfo {volume} {475}},\ \bibinfo
  {pages} {2971} (\bibinfo {year} {2018})},\ \Eprint
  {http://arxiv.org/abs/1710.05897} {arXiv:1710.05897 [astro-ph.HE]}
  \BibitemShut {NoStop}%
\bibitem [{\citenamefont {Piro}\ and\ \citenamefont
  {Kollmeier}(2017)}]{Piro:2017ayh}%
  \BibitemOpen
  \bibfield  {author} {\bibinfo {author} {\bibfnamefont {A.~L.}\ \bibnamefont
  {Piro}}\ and\ \bibinfo {author} {\bibfnamefont {J.~A.}\ \bibnamefont
  {Kollmeier}},\ }\href@noop {} {\  (\bibinfo {year} {2017})},\ \Eprint
  {http://arxiv.org/abs/1710.05822} {arXiv:1710.05822 [astro-ph.HE]}
  \BibitemShut {NoStop}%
\bibitem [{\citenamefont {Duffell}\ \emph {et~al.}(2018)\citenamefont
  {Duffell}, \citenamefont {Quataert}, \citenamefont {Kasen},\ and\
  \citenamefont {Klion}}]{Duffell:2018iig}%
  \BibitemOpen
  \bibfield  {author} {\bibinfo {author} {\bibfnamefont {P.~C.}\ \bibnamefont
  {Duffell}}, \bibinfo {author} {\bibfnamefont {E.}~\bibnamefont {Quataert}},
  \bibinfo {author} {\bibfnamefont {D.}~\bibnamefont {Kasen}}, \ and\ \bibinfo
  {author} {\bibfnamefont {H.}~\bibnamefont {Klion}},\ }\href {\doibase
  10.3847/1538-4357/aae084} {\bibfield  {journal} {\bibinfo  {journal}
  {Astrophys. J.}\ }\textbf {\bibinfo {volume} {866}},\ \bibinfo {pages} {3}
  (\bibinfo {year} {2018})},\ \Eprint {http://arxiv.org/abs/1806.10616}
  {arXiv:1806.10616 [astro-ph.HE]} \BibitemShut {NoStop}%
\bibitem [{\citenamefont {Metzger}\ \emph {et~al.}(2018)\citenamefont
  {Metzger}, \citenamefont {Thompson},\ and\ \citenamefont
  {Quataert}}]{Metzger:2018uni}%
  \BibitemOpen
  \bibfield  {author} {\bibinfo {author} {\bibfnamefont {B.~D.}\ \bibnamefont
  {Metzger}}, \bibinfo {author} {\bibfnamefont {T.~A.}\ \bibnamefont
  {Thompson}}, \ and\ \bibinfo {author} {\bibfnamefont {E.}~\bibnamefont
  {Quataert}},\ }\href {\doibase 10.3847/1538-4357/aab095} {\bibfield
  {journal} {\bibinfo  {journal} {Astrophys. J.}\ }\textbf {\bibinfo {volume}
  {856}},\ \bibinfo {pages} {101} (\bibinfo {year} {2018})},\ \Eprint
  {http://arxiv.org/abs/1801.04286} {arXiv:1801.04286 [astro-ph.HE]}
  \BibitemShut {NoStop}%
\bibitem [{\citenamefont {Radice}\ \emph
  {et~al.}(2018{\natexlab{d}})\citenamefont {Radice}, \citenamefont {Perego},
  \citenamefont {Hotokezaka}, \citenamefont {Bernuzzi}, \citenamefont {Fromm},\
  and\ \citenamefont {Roberts}}]{Radice:2018ghv}%
  \BibitemOpen
  \bibfield  {author} {\bibinfo {author} {\bibfnamefont {D.}~\bibnamefont
  {Radice}}, \bibinfo {author} {\bibfnamefont {A.}~\bibnamefont {Perego}},
  \bibinfo {author} {\bibfnamefont {K.}~\bibnamefont {Hotokezaka}}, \bibinfo
  {author} {\bibfnamefont {S.}~\bibnamefont {Bernuzzi}}, \bibinfo {author}
  {\bibfnamefont {S.~A.}\ \bibnamefont {Fromm}}, \ and\ \bibinfo {author}
  {\bibfnamefont {L.~F.}\ \bibnamefont {Roberts}},\ }\href {\doibase
  10.3847/2041-8213/aaf053} {\bibfield  {journal} {\bibinfo  {journal}
  {Astrophys. J. Lett.}\ }\textbf {\bibinfo {volume} {869}},\ \bibinfo {pages}
  {L35} (\bibinfo {year} {2018}{\natexlab{d}})},\ \Eprint
  {http://arxiv.org/abs/1809.11163} {arXiv:1809.11163 [astro-ph.HE]}
  \BibitemShut {NoStop}%
\bibitem [{\citenamefont {Typel}\ \emph {et~al.}(2010)\citenamefont {Typel},
  \citenamefont {Ropke}, \citenamefont {Klahn}, \citenamefont {Blaschke},\ and\
  \citenamefont {Wolter}}]{Typel:2009sy}%
  \BibitemOpen
  \bibfield  {author} {\bibinfo {author} {\bibfnamefont {S.}~\bibnamefont
  {Typel}}, \bibinfo {author} {\bibfnamefont {G.}~\bibnamefont {Ropke}},
  \bibinfo {author} {\bibfnamefont {T.}~\bibnamefont {Klahn}}, \bibinfo
  {author} {\bibfnamefont {D.}~\bibnamefont {Blaschke}}, \ and\ \bibinfo
  {author} {\bibfnamefont {H.~H.}\ \bibnamefont {Wolter}},\ }\href {\doibase
  10.1103/PhysRevC.81.015803} {\bibfield  {journal} {\bibinfo  {journal} {Phys.
  Rev.}\ }\textbf {\bibinfo {volume} {C81}},\ \bibinfo {pages} {015803}
  (\bibinfo {year} {2010})},\ \Eprint {http://arxiv.org/abs/0908.2344}
  {arXiv:0908.2344 [nucl-th]} \BibitemShut {NoStop}%
\bibitem [{\citenamefont {Hempel}\ and\ \citenamefont
  {Schaffner-Bielich}(2010)}]{Hempel:2009mc}%
  \BibitemOpen
  \bibfield  {author} {\bibinfo {author} {\bibfnamefont {M.}~\bibnamefont
  {Hempel}}\ and\ \bibinfo {author} {\bibfnamefont {J.}~\bibnamefont
  {Schaffner-Bielich}},\ }\href {\doibase 10.1016/j.nuclphysa.2010.02.010}
  {\bibfield  {journal} {\bibinfo  {journal} {Nucl. Phys.}\ }\textbf {\bibinfo
  {volume} {A837}},\ \bibinfo {pages} {210} (\bibinfo {year} {2010})},\ \Eprint
  {http://arxiv.org/abs/0911.4073} {arXiv:0911.4073 [nucl-th]} \BibitemShut
  {NoStop}%
\bibitem [{\citenamefont {Lattimer}\ and\ \citenamefont
  {Swesty}(1991)}]{Lattimer:1991nc}%
  \BibitemOpen
  \bibfield  {author} {\bibinfo {author} {\bibfnamefont {J.~M.}\ \bibnamefont
  {Lattimer}}\ and\ \bibinfo {author} {\bibfnamefont {F.~D.}\ \bibnamefont
  {Swesty}},\ }\href {\doibase 10.1016/0375-9474(91)90452-C} {\bibfield
  {journal} {\bibinfo  {journal} {Nucl. Phys.}\ }\textbf {\bibinfo {volume}
  {A535}},\ \bibinfo {pages} {331} (\bibinfo {year} {1991})}\BibitemShut
  {NoStop}%
\bibitem [{\citenamefont {Radice}\ and\ \citenamefont
  {Rezzolla}(2012)}]{Radice:2012cu}%
  \BibitemOpen
  \bibfield  {author} {\bibinfo {author} {\bibfnamefont {D.}~\bibnamefont
  {Radice}}\ and\ \bibinfo {author} {\bibfnamefont {L.}~\bibnamefont
  {Rezzolla}},\ }\href {\doibase 10.1051/0004-6361/201219735} {\bibfield
  {journal} {\bibinfo  {journal} {Astron. Astrophys.}\ }\textbf {\bibinfo
  {volume} {547}},\ \bibinfo {pages} {A26} (\bibinfo {year} {2012})},\ \Eprint
  {http://arxiv.org/abs/1206.6502} {arXiv:1206.6502 [astro-ph.IM]} \BibitemShut
  {NoStop}%
\bibitem [{\citenamefont {Radice}\ \emph
  {et~al.}(2014{\natexlab{a}})\citenamefont {Radice}, \citenamefont
  {Rezzolla},\ and\ \citenamefont {Galeazzi}}]{Radice:2013hxh}%
  \BibitemOpen
  \bibfield  {author} {\bibinfo {author} {\bibfnamefont {D.}~\bibnamefont
  {Radice}}, \bibinfo {author} {\bibfnamefont {L.}~\bibnamefont {Rezzolla}}, \
  and\ \bibinfo {author} {\bibfnamefont {F.}~\bibnamefont {Galeazzi}},\ }\href
  {\doibase 10.1093/mnrasl/slt137} {\bibfield  {journal} {\bibinfo  {journal}
  {Mon.Not.Roy.Astron.Soc.}\ }\textbf {\bibinfo {volume} {437}},\ \bibinfo
  {pages} {L46} (\bibinfo {year} {2014}{\natexlab{a}})},\ \Eprint
  {http://arxiv.org/abs/1306.6052} {arXiv:1306.6052 [gr-qc]} \BibitemShut
  {NoStop}%
\bibitem [{\citenamefont {Radice}\ \emph
  {et~al.}(2014{\natexlab{b}})\citenamefont {Radice}, \citenamefont
  {Rezzolla},\ and\ \citenamefont {Galeazzi}}]{Radice:2013xpa}%
  \BibitemOpen
  \bibfield  {author} {\bibinfo {author} {\bibfnamefont {D.}~\bibnamefont
  {Radice}}, \bibinfo {author} {\bibfnamefont {L.}~\bibnamefont {Rezzolla}}, \
  and\ \bibinfo {author} {\bibfnamefont {F.}~\bibnamefont {Galeazzi}},\ }\href
  {\doibase 10.1088/0264-9381/31/7/075012} {\bibfield  {journal} {\bibinfo
  {journal} {Class.Quant.Grav.}\ }\textbf {\bibinfo {volume} {31}},\ \bibinfo
  {pages} {075012} (\bibinfo {year} {2014}{\natexlab{b}})},\ \Eprint
  {http://arxiv.org/abs/1312.5004} {arXiv:1312.5004 [gr-qc]} \BibitemShut
  {NoStop}%
\bibitem [{\citenamefont {Radice}\ \emph
  {et~al.}(2016{\natexlab{a}})\citenamefont {Radice}, \citenamefont {Galeazzi},
  \citenamefont {Lippuner}, \citenamefont {Roberts}, \citenamefont {Ott},\ and\
  \citenamefont {Rezzolla}}]{Radice:2016dwd}%
  \BibitemOpen
  \bibfield  {author} {\bibinfo {author} {\bibfnamefont {D.}~\bibnamefont
  {Radice}}, \bibinfo {author} {\bibfnamefont {F.}~\bibnamefont {Galeazzi}},
  \bibinfo {author} {\bibfnamefont {J.}~\bibnamefont {Lippuner}}, \bibinfo
  {author} {\bibfnamefont {L.~F.}\ \bibnamefont {Roberts}}, \bibinfo {author}
  {\bibfnamefont {C.~D.}\ \bibnamefont {Ott}}, \ and\ \bibinfo {author}
  {\bibfnamefont {L.}~\bibnamefont {Rezzolla}},\ }\href {\doibase
  10.1093/mnras/stw1227} {\bibfield  {journal} {\bibinfo  {journal} {Mon. Not.
  Roy. Astron. Soc.}\ }\textbf {\bibinfo {volume} {460}},\ \bibinfo {pages}
  {3255} (\bibinfo {year} {2016}{\natexlab{a}})},\ \Eprint
  {http://arxiv.org/abs/1601.02426} {arXiv:1601.02426 [astro-ph.HE]}
  \BibitemShut {NoStop}%
\bibitem [{\citenamefont {Radice}(2017)}]{Radice:2017zta}%
  \BibitemOpen
  \bibfield  {author} {\bibinfo {author} {\bibfnamefont {D.}~\bibnamefont
  {Radice}},\ }\href {\doibase 10.3847/2041-8213/aa6483} {\bibfield  {journal}
  {\bibinfo  {journal} {Astrophys. J.}\ }\textbf {\bibinfo {volume} {838}},\
  \bibinfo {pages} {L2} (\bibinfo {year} {2017})},\ \Eprint
  {http://arxiv.org/abs/1703.02046} {arXiv:1703.02046 [astro-ph.HE]}
  \BibitemShut {NoStop}%
\bibitem [{\citenamefont {Ruffert}\ \emph {et~al.}(1996)\citenamefont
  {Ruffert}, \citenamefont {Janka},\ and\ \citenamefont
  {Sch{\"a}fer}}]{Ruffert:1995fs}%
  \BibitemOpen
  \bibfield  {author} {\bibinfo {author} {\bibfnamefont {M.~H.}\ \bibnamefont
  {Ruffert}}, \bibinfo {author} {\bibfnamefont {H.~T.}\ \bibnamefont {Janka}},
  \ and\ \bibinfo {author} {\bibfnamefont {G.}~\bibnamefont {Sch{\"a}fer}},\
  }\href@noop {} {\bibfield  {journal} {\bibinfo  {journal} {Astron.
  Astrophys.}\ }\textbf {\bibinfo {volume} {311}},\ \bibinfo {pages} {532}
  (\bibinfo {year} {1996})},\ \Eprint {http://arxiv.org/abs/astro-ph/9509006}
  {arXiv:astro-ph/9509006} \BibitemShut {NoStop}%
\bibitem [{\citenamefont {Neilsen}\ \emph {et~al.}(2014)\citenamefont
  {Neilsen}, \citenamefont {Liebling}, \citenamefont {Anderson}, \citenamefont
  {Lehner}, \citenamefont {O’Connor} \emph {et~al.}}]{Neilsen:2014hha}%
  \BibitemOpen
  \bibfield  {author} {\bibinfo {author} {\bibfnamefont {D.}~\bibnamefont
  {Neilsen}}, \bibinfo {author} {\bibfnamefont {S.~L.}\ \bibnamefont
  {Liebling}}, \bibinfo {author} {\bibfnamefont {M.}~\bibnamefont {Anderson}},
  \bibinfo {author} {\bibfnamefont {L.}~\bibnamefont {Lehner}}, \bibinfo
  {author} {\bibfnamefont {E.}~\bibnamefont {O’Connor}},  \emph {et~al.},\
  }\href {\doibase 10.1103/PhysRevD.89.104029} {\bibfield  {journal} {\bibinfo
  {journal} {Phys.Rev.}\ }\textbf {\bibinfo {volume} {D89}},\ \bibinfo {pages}
  {104029} (\bibinfo {year} {2014})},\ \Eprint {http://arxiv.org/abs/1403.3680}
  {arXiv:1403.3680 [gr-qc]} \BibitemShut {NoStop}%
\bibitem [{\citenamefont {Kiuchi}\ \emph {et~al.}(2018)\citenamefont {Kiuchi},
  \citenamefont {Kyutoku}, \citenamefont {Sekiguchi},\ and\ \citenamefont
  {Shibata}}]{Kiuchi:2017zzg}%
  \BibitemOpen
  \bibfield  {author} {\bibinfo {author} {\bibfnamefont {K.}~\bibnamefont
  {Kiuchi}}, \bibinfo {author} {\bibfnamefont {K.}~\bibnamefont {Kyutoku}},
  \bibinfo {author} {\bibfnamefont {Y.}~\bibnamefont {Sekiguchi}}, \ and\
  \bibinfo {author} {\bibfnamefont {M.}~\bibnamefont {Shibata}},\ }\href
  {\doibase 10.1103/PhysRevD.97.124039} {\bibfield  {journal} {\bibinfo
  {journal} {Phys. Rev.}\ }\textbf {\bibinfo {volume} {D97}},\ \bibinfo {pages}
  {124039} (\bibinfo {year} {2018})},\ \Eprint
  {http://arxiv.org/abs/1710.01311} {arXiv:1710.01311 [astro-ph.HE]}
  \BibitemShut {NoStop}%
\bibitem [{\citenamefont {Kastaun}\ and\ \citenamefont
  {Galeazzi}(2015)}]{Kastaun:2014fna}%
  \BibitemOpen
  \bibfield  {author} {\bibinfo {author} {\bibfnamefont {W.}~\bibnamefont
  {Kastaun}}\ and\ \bibinfo {author} {\bibfnamefont {F.}~\bibnamefont
  {Galeazzi}},\ }\href {\doibase 10.1103/PhysRevD.91.064027} {\bibfield
  {journal} {\bibinfo  {journal} {Phys.Rev.}\ }\textbf {\bibinfo {volume}
  {D91}},\ \bibinfo {pages} {064027} (\bibinfo {year} {2015})},\ \Eprint
  {http://arxiv.org/abs/1411.7975} {arXiv:1411.7975 [gr-qc]} \BibitemShut
  {NoStop}%
\bibitem [{\citenamefont {Shibata}\ and\ \citenamefont
  {Uryu}(2000)}]{Shibata:1999wm}%
  \BibitemOpen
  \bibfield  {author} {\bibinfo {author} {\bibfnamefont {M.}~\bibnamefont
  {Shibata}}\ and\ \bibinfo {author} {\bibfnamefont {K.}~\bibnamefont {Uryu}},\
  }\href {\doibase 10.1103/PhysRevD.61.064001} {\bibfield  {journal} {\bibinfo
  {journal} {Phys. Rev.}\ }\textbf {\bibinfo {volume} {D61}},\ \bibinfo {pages}
  {064001} (\bibinfo {year} {2000})},\ \Eprint
  {http://arxiv.org/abs/gr-qc/9911058} {arXiv:gr-qc/9911058} \BibitemShut
  {NoStop}%
\bibitem [{\citenamefont {Shibata}\ and\ \citenamefont
  {Taniguchi}(2006)}]{Shibata:2006nm}%
  \BibitemOpen
  \bibfield  {author} {\bibinfo {author} {\bibfnamefont {M.}~\bibnamefont
  {Shibata}}\ and\ \bibinfo {author} {\bibfnamefont {K.}~\bibnamefont
  {Taniguchi}},\ }\href {\doibase 10.1103/PhysRevD.73.064027} {\bibfield
  {journal} {\bibinfo  {journal} {Phys.Rev.}\ }\textbf {\bibinfo {volume}
  {D73}},\ \bibinfo {pages} {064027} (\bibinfo {year} {2006})},\ \Eprint
  {http://arxiv.org/abs/astro-ph/0603145} {arXiv:astro-ph/0603145 [astro-ph]}
  \BibitemShut {NoStop}%
\bibitem [{\citenamefont {Bernuzzi}\ \emph {et~al.}(2014)\citenamefont
  {Bernuzzi}, \citenamefont {Dietrich}, \citenamefont {Tichy},\ and\
  \citenamefont {Br{\"u}gmann}}]{Bernuzzi:2013rza}%
  \BibitemOpen
  \bibfield  {author} {\bibinfo {author} {\bibfnamefont {S.}~\bibnamefont
  {Bernuzzi}}, \bibinfo {author} {\bibfnamefont {T.}~\bibnamefont {Dietrich}},
  \bibinfo {author} {\bibfnamefont {W.}~\bibnamefont {Tichy}}, \ and\ \bibinfo
  {author} {\bibfnamefont {B.}~\bibnamefont {Br{\"u}gmann}},\ }\href {\doibase
  10.1103/PhysRevD.89.104021} {\bibfield  {journal} {\bibinfo  {journal}
  {Phys.Rev.}\ }\textbf {\bibinfo {volume} {D89}},\ \bibinfo {pages} {104021}
  (\bibinfo {year} {2014})},\ \Eprint {http://arxiv.org/abs/1311.4443}
  {arXiv:1311.4443 [gr-qc]} \BibitemShut {NoStop}%
\bibitem [{\citenamefont {Bernuzzi}\ \emph {et~al.}(2016)\citenamefont
  {Bernuzzi}, \citenamefont {Radice}, \citenamefont {Ott}, \citenamefont
  {Roberts}, \citenamefont {Moesta},\ and\ \citenamefont
  {Galeazzi}}]{Bernuzzi:2015opx}%
  \BibitemOpen
  \bibfield  {author} {\bibinfo {author} {\bibfnamefont {S.}~\bibnamefont
  {Bernuzzi}}, \bibinfo {author} {\bibfnamefont {D.}~\bibnamefont {Radice}},
  \bibinfo {author} {\bibfnamefont {C.~D.}\ \bibnamefont {Ott}}, \bibinfo
  {author} {\bibfnamefont {L.~F.}\ \bibnamefont {Roberts}}, \bibinfo {author}
  {\bibfnamefont {P.}~\bibnamefont {Moesta}}, \ and\ \bibinfo {author}
  {\bibfnamefont {F.}~\bibnamefont {Galeazzi}},\ }\href {\doibase
  10.1103/PhysRevD.94.024023} {\bibfield  {journal} {\bibinfo  {journal} {Phys.
  Rev.}\ }\textbf {\bibinfo {volume} {D94}},\ \bibinfo {pages} {024023}
  (\bibinfo {year} {2016})},\ \Eprint {http://arxiv.org/abs/1512.06397}
  {arXiv:1512.06397 [gr-qc]} \BibitemShut {NoStop}%
\bibitem [{\citenamefont {East}\ \emph {et~al.}(2016)\citenamefont {East},
  \citenamefont {Paschalidis}, \citenamefont {Pretorius},\ and\ \citenamefont
  {Shapiro}}]{East:2015vix}%
  \BibitemOpen
  \bibfield  {author} {\bibinfo {author} {\bibfnamefont {W.~E.}\ \bibnamefont
  {East}}, \bibinfo {author} {\bibfnamefont {V.}~\bibnamefont {Paschalidis}},
  \bibinfo {author} {\bibfnamefont {F.}~\bibnamefont {Pretorius}}, \ and\
  \bibinfo {author} {\bibfnamefont {S.~L.}\ \bibnamefont {Shapiro}},\ }\href
  {\doibase 10.1103/PhysRevD.93.024011} {\bibfield  {journal} {\bibinfo
  {journal} {Phys. Rev.}\ }\textbf {\bibinfo {volume} {D93}},\ \bibinfo {pages}
  {024011} (\bibinfo {year} {2016})},\ \Eprint
  {http://arxiv.org/abs/1511.01093} {arXiv:1511.01093 [astro-ph.HE]}
  \BibitemShut {NoStop}%
\bibitem [{\citenamefont {Paschalidis}\ \emph {et~al.}(2015)\citenamefont
  {Paschalidis}, \citenamefont {East}, \citenamefont {Pretorius},\ and\
  \citenamefont {Shapiro}}]{Paschalidis:2015mla}%
  \BibitemOpen
  \bibfield  {author} {\bibinfo {author} {\bibfnamefont {V.}~\bibnamefont
  {Paschalidis}}, \bibinfo {author} {\bibfnamefont {W.~E.}\ \bibnamefont
  {East}}, \bibinfo {author} {\bibfnamefont {F.}~\bibnamefont {Pretorius}}, \
  and\ \bibinfo {author} {\bibfnamefont {S.~L.}\ \bibnamefont {Shapiro}},\
  }\href {\doibase 10.1103/PhysRevD.92.121502} {\bibfield  {journal} {\bibinfo
  {journal} {Phys. Rev.}\ }\textbf {\bibinfo {volume} {D92}},\ \bibinfo {pages}
  {121502} (\bibinfo {year} {2015})},\ \Eprint
  {http://arxiv.org/abs/1510.03432} {arXiv:1510.03432 [astro-ph.HE]}
  \BibitemShut {NoStop}%
\bibitem [{\citenamefont {Radice}\ \emph
  {et~al.}(2016{\natexlab{b}})\citenamefont {Radice}, \citenamefont
  {Bernuzzi},\ and\ \citenamefont {Ott}}]{Radice:2016gym}%
  \BibitemOpen
  \bibfield  {author} {\bibinfo {author} {\bibfnamefont {D.}~\bibnamefont
  {Radice}}, \bibinfo {author} {\bibfnamefont {S.}~\bibnamefont {Bernuzzi}}, \
  and\ \bibinfo {author} {\bibfnamefont {C.~D.}\ \bibnamefont {Ott}},\ }\href
  {\doibase 10.1103/PhysRevD.94.064011} {\bibfield  {journal} {\bibinfo
  {journal} {Phys. Rev.}\ }\textbf {\bibinfo {volume} {D94}},\ \bibinfo {pages}
  {064011} (\bibinfo {year} {2016}{\natexlab{b}})},\ \Eprint
  {http://arxiv.org/abs/1603.05726} {arXiv:1603.05726 [gr-qc]} \BibitemShut
  {NoStop}%
\bibitem [{\citenamefont {Lehner}\ \emph {et~al.}(2016)\citenamefont {Lehner},
  \citenamefont {Liebling}, \citenamefont {Palenzuela},\ and\ \citenamefont
  {Motl}}]{Lehner:2016wjg}%
  \BibitemOpen
  \bibfield  {author} {\bibinfo {author} {\bibfnamefont {L.}~\bibnamefont
  {Lehner}}, \bibinfo {author} {\bibfnamefont {S.~L.}\ \bibnamefont
  {Liebling}}, \bibinfo {author} {\bibfnamefont {C.}~\bibnamefont
  {Palenzuela}}, \ and\ \bibinfo {author} {\bibfnamefont {P.~M.}\ \bibnamefont
  {Motl}},\ }\href {\doibase 10.1103/PhysRevD.94.043003} {\bibfield  {journal}
  {\bibinfo  {journal} {Phys. Rev.}\ }\textbf {\bibinfo {volume} {D94}},\
  \bibinfo {pages} {043003} (\bibinfo {year} {2016})},\ \Eprint
  {http://arxiv.org/abs/1605.02369} {arXiv:1605.02369 [gr-qc]} \BibitemShut
  {NoStop}%
\bibitem [{\citenamefont {{Goodman}}\ and\ \citenamefont
  {{Rafikov}}(2001)}]{Goodman:2001a}%
  \BibitemOpen
  \bibfield  {author} {\bibinfo {author} {\bibfnamefont {J.}~\bibnamefont
  {{Goodman}}}\ and\ \bibinfo {author} {\bibfnamefont {R.~R.}\ \bibnamefont
  {{Rafikov}}},\ }\href {\doibase 10.1086/320572} {\bibfield  {journal}
  {\bibinfo  {journal} {Astrophys.\ J.}\ }\textbf {\bibinfo {volume} {552}},\
  \bibinfo {pages} {793} (\bibinfo {year} {2001})},\ \Eprint
  {http://arxiv.org/abs/astro-ph/0010576} {astro-ph/0010576} \BibitemShut
  {NoStop}%
\bibitem [{\citenamefont {{Rafikov}}(2016)}]{Rafikov:2016a}%
  \BibitemOpen
  \bibfield  {author} {\bibinfo {author} {\bibfnamefont {R.~R.}\ \bibnamefont
  {{Rafikov}}},\ }\href {\doibase 10.3847/0004-637X/831/2/122} {\bibfield
  {journal} {\bibinfo  {journal} {Astrophys.\ J.}\ }\textbf {\bibinfo {volume}
  {831}},\ \bibinfo {eid} {122} (\bibinfo {year} {2016})},\ \Eprint
  {http://arxiv.org/abs/1601.03009} {arXiv:1601.03009 [astro-ph.EP]}
  \BibitemShut {NoStop}%
\bibitem [{\citenamefont {{Arzamasskiy}}\ and\ \citenamefont
  {{Rafikov}}(2018)}]{Arzamasskiy:2018a}%
  \BibitemOpen
  \bibfield  {author} {\bibinfo {author} {\bibfnamefont {L.}~\bibnamefont
  {{Arzamasskiy}}}\ and\ \bibinfo {author} {\bibfnamefont {R.~R.}\ \bibnamefont
  {{Rafikov}}},\ }\href {\doibase 10.3847/1538-4357/aaa8e8} {\bibfield
  {journal} {\bibinfo  {journal} {Astrophys.\ J.}\ }\textbf {\bibinfo {volume}
  {854}},\ \bibinfo {eid} {84} (\bibinfo {year} {2018})},\ \Eprint
  {http://arxiv.org/abs/1710.01304} {arXiv:1710.01304 [astro-ph.EP]}
  \BibitemShut {NoStop}%
\bibitem [{\citenamefont {Arlandini}\ \emph {et~al.}(1999)\citenamefont
  {Arlandini}, \citenamefont {Kaeppeler}, \citenamefont {Wisshak},
  \citenamefont {Gallino}, \citenamefont {Lugaro}, \citenamefont {Busso},\ and\
  \citenamefont {Straniero}}]{Arlandini:1999an}%
  \BibitemOpen
  \bibfield  {author} {\bibinfo {author} {\bibfnamefont {C.}~\bibnamefont
  {Arlandini}}, \bibinfo {author} {\bibfnamefont {F.}~\bibnamefont
  {Kaeppeler}}, \bibinfo {author} {\bibfnamefont {K.}~\bibnamefont {Wisshak}},
  \bibinfo {author} {\bibfnamefont {R.}~\bibnamefont {Gallino}}, \bibinfo
  {author} {\bibfnamefont {M.}~\bibnamefont {Lugaro}}, \bibinfo {author}
  {\bibfnamefont {M.}~\bibnamefont {Busso}}, \ and\ \bibinfo {author}
  {\bibfnamefont {O.}~\bibnamefont {Straniero}},\ }\href {\doibase
  10.1086/307938} {\bibfield  {journal} {\bibinfo  {journal} {Astrophys. J.}\
  }\textbf {\bibinfo {volume} {525}},\ \bibinfo {pages} {886} (\bibinfo {year}
  {1999})},\ \Eprint {http://arxiv.org/abs/astro-ph/9906266}
  {arXiv:astro-ph/9906266 [astro-ph]} \BibitemShut {NoStop}%
\bibitem [{\citenamefont {Smartt}\ \emph {et~al.}(2017)\citenamefont {Smartt}
  \emph {et~al.}}]{Smartt:2017fuw}%
  \BibitemOpen
  \bibfield  {author} {\bibinfo {author} {\bibfnamefont {S.~J.}\ \bibnamefont
  {Smartt}} \emph {et~al.},\ }\href {\doibase 10.1038/nature24303} {\bibfield
  {journal} {\bibinfo  {journal} {Nature}\ } (\bibinfo {year} {2017}),\
  10.1038/nature24303},\ \Eprint {http://arxiv.org/abs/1710.05841}
  {arXiv:1710.05841 [astro-ph.HE]} \BibitemShut {NoStop}%
\bibitem [{\citenamefont {Kasen}\ \emph {et~al.}(2017)\citenamefont {Kasen},
  \citenamefont {Metzger}, \citenamefont {Barnes}, \citenamefont {Quataert},\
  and\ \citenamefont {Ramirez-Ruiz}}]{Kasen:2017sxr}%
  \BibitemOpen
  \bibfield  {author} {\bibinfo {author} {\bibfnamefont {D.}~\bibnamefont
  {Kasen}}, \bibinfo {author} {\bibfnamefont {B.}~\bibnamefont {Metzger}},
  \bibinfo {author} {\bibfnamefont {J.}~\bibnamefont {Barnes}}, \bibinfo
  {author} {\bibfnamefont {E.}~\bibnamefont {Quataert}}, \ and\ \bibinfo
  {author} {\bibfnamefont {E.}~\bibnamefont {Ramirez-Ruiz}},\ }\href {\doibase
  10.1038/nature24453} {\bibfield  {journal} {\bibinfo  {journal} {Nature}\ }
  (\bibinfo {year} {2017}),\ 10.1038/nature24453},\ \bibinfo {note}
  {[Nature551,80(2017)]},\ \Eprint {http://arxiv.org/abs/1710.05463}
  {arXiv:1710.05463 [astro-ph.HE]} \BibitemShut {NoStop}%
\bibitem [{\citenamefont {Bulla}(2019)}]{Bulla:2019muo}%
  \BibitemOpen
  \bibfield  {author} {\bibinfo {author} {\bibfnamefont {M.}~\bibnamefont
  {Bulla}},\ }\href {\doibase 10.1093/mnras/stz2495} {\bibfield  {journal}
  {\bibinfo  {journal} {Mon. Not. Roy. Astron. Soc.}\ }\textbf {\bibinfo
  {volume} {489}},\ \bibinfo {pages} {5037} (\bibinfo {year} {2019})},\ \Eprint
  {http://arxiv.org/abs/1906.04205} {arXiv:1906.04205 [astro-ph.HE]}
  \BibitemShut {NoStop}%
\bibitem [{\citenamefont {Eichler}\ \emph {et~al.}(2015)\citenamefont {Eichler}
  \emph {et~al.}}]{Eichler:2014kma}%
  \BibitemOpen
  \bibfield  {author} {\bibinfo {author} {\bibfnamefont {M.}~\bibnamefont
  {Eichler}} \emph {et~al.},\ }\href {\doibase 10.1088/0004-637X/808/1/30}
  {\bibfield  {journal} {\bibinfo  {journal} {Astrophys. J.}\ }\textbf
  {\bibinfo {volume} {808}},\ \bibinfo {pages} {30} (\bibinfo {year} {2015})},\
  \Eprint {http://arxiv.org/abs/1411.0974} {arXiv:1411.0974 [astro-ph.HE]}
  \BibitemShut {NoStop}%
\bibitem [{\citenamefont {Rosswog}\ \emph {et~al.}(2017)\citenamefont
  {Rosswog}, \citenamefont {Feindt}, \citenamefont {Korobkin}, \citenamefont
  {Wu}, \citenamefont {Sollerman}, \citenamefont {Goobar},\ and\ \citenamefont
  {Martinez-Pinedo}}]{Rosswog:2016dhy}%
  \BibitemOpen
  \bibfield  {author} {\bibinfo {author} {\bibfnamefont {S.}~\bibnamefont
  {Rosswog}}, \bibinfo {author} {\bibfnamefont {U.}~\bibnamefont {Feindt}},
  \bibinfo {author} {\bibfnamefont {O.}~\bibnamefont {Korobkin}}, \bibinfo
  {author} {\bibfnamefont {M.~R.}\ \bibnamefont {Wu}}, \bibinfo {author}
  {\bibfnamefont {J.}~\bibnamefont {Sollerman}}, \bibinfo {author}
  {\bibfnamefont {A.}~\bibnamefont {Goobar}}, \ and\ \bibinfo {author}
  {\bibfnamefont {G.}~\bibnamefont {Martinez-Pinedo}},\ }\href {\doibase
  10.1088/1361-6382/aa68a9} {\bibfield  {journal} {\bibinfo  {journal} {Class.
  Quant. Grav.}\ }\textbf {\bibinfo {volume} {34}},\ \bibinfo {pages} {104001}
  (\bibinfo {year} {2017})},\ \Eprint {http://arxiv.org/abs/1611.09822}
  {arXiv:1611.09822 [astro-ph.HE]} \BibitemShut {NoStop}%
\bibitem [{\citenamefont {Gaigalas}\ \emph {et~al.}(2019)\citenamefont
  {Gaigalas}, \citenamefont {Kato}, \citenamefont {Rynkun}, \citenamefont
  {Radziute},\ and\ \citenamefont {Tanaka}}]{Gaigalas:2019ptx}%
  \BibitemOpen
  \bibfield  {author} {\bibinfo {author} {\bibfnamefont {G.}~\bibnamefont
  {Gaigalas}}, \bibinfo {author} {\bibfnamefont {D.}~\bibnamefont {Kato}},
  \bibinfo {author} {\bibfnamefont {P.}~\bibnamefont {Rynkun}}, \bibinfo
  {author} {\bibfnamefont {L.}~\bibnamefont {Radziute}}, \ and\ \bibinfo
  {author} {\bibfnamefont {M.}~\bibnamefont {Tanaka}},\ }\href {\doibase
  10.3847/1538-4365/aaf9b8} {\bibfield  {journal} {\bibinfo  {journal}
  {Astrophys. J. Suppl.}\ }\textbf {\bibinfo {volume} {240}},\ \bibinfo {pages}
  {29} (\bibinfo {year} {2019})},\ \Eprint {http://arxiv.org/abs/1901.10671}
  {arXiv:1901.10671 [astro-ph.SR]} \BibitemShut {NoStop}%
\bibitem [{\citenamefont {Ciolfi}\ \emph {et~al.}(2019)\citenamefont {Ciolfi},
  \citenamefont {Kastaun}, \citenamefont {Kalinani},\ and\ \citenamefont
  {Giacomazzo}}]{Ciolfi:2019fie}%
  \BibitemOpen
  \bibfield  {author} {\bibinfo {author} {\bibfnamefont {R.}~\bibnamefont
  {Ciolfi}}, \bibinfo {author} {\bibfnamefont {W.}~\bibnamefont {Kastaun}},
  \bibinfo {author} {\bibfnamefont {J.~V.}\ \bibnamefont {Kalinani}}, \ and\
  \bibinfo {author} {\bibfnamefont {B.}~\bibnamefont {Giacomazzo}},\ }\href
  {\doibase 10.1103/PhysRevD.100.023005} {\bibfield  {journal} {\bibinfo
  {journal} {Phys. Rev.}\ }\textbf {\bibinfo {volume} {D100}},\ \bibinfo
  {pages} {023005} (\bibinfo {year} {2019})},\ \Eprint
  {http://arxiv.org/abs/1904.10222} {arXiv:1904.10222 [astro-ph.HE]}
  \BibitemShut {NoStop}%
\end{thebibliography}%

\end{document}